\documentclass{aa}

\usepackage{graphicx}
\usepackage{txfonts}
\usepackage[colorlinks=true,linkcolor=blue,citecolor=blue,urlcolor=blue]{hyperref}
\usepackage{float}
\usepackage{physics}
\usepackage{amsmath}

\usepackage[dvipsnames]{xcolor}
\usepackage{soul}

\begin{document}

   \title{Tidal Dissipation in Evolved Low and Intermediate Mass Stars}

   \author{M. Esseldeurs\inst{1}
          \and
          S. Mathis\inst{2}
          \and
          L. Decin\inst{1}
          }

   \institute{Instituut voor Sterrenkunde, KU Leuven, Celestijnenlaan 200D, 3001 Leuven, Belgium \\
              \email{mats.esseldeurs@kuleuven.be}
         \and
             Université Paris-Saclay, Université Paris Cité, CEA, CNRS, AIM, 91191 Gif-sur-Yvette, France
             }

   \date{Received 17 February 2024 / Accepted 12 July 2024}
 
  \abstract
   {As the observed occurrence for planets or stellar companions orbiting low and intermediate-mass evolved stars is increasing, so does the importance of understanding and evaluating the strength of their interactions. This is important for both the further evolution of our own Earth-Sun system, as well as most of the observed exoplanetary systems. One of the most fundamental mechanisms to understand this interaction is the tidal dissipation in these stars, as it is one of the engines of orbital/rotational evolution of star-planet/star-star systems.}
   {This article builds on previous works studying the evolution of the tidal dissipation along the pre-main-sequence and the main-sequence of low-mass and intermediate-mass stars, which have shown the strong link between the structural and rotational evolution of stars and tidal dissipation. This article provides for the first time a complete picture of tidal dissipation along the entire evolution of low and intermediate-mass stars, including the advanced phases of evolution.}
   {Using stellar evolutionary models, the internal structure of the star is computed from the pre-main sequence all the way up to the white dwarf phase, and this for stars with initial mass between 1 and 4 M$_\odot$. Using this internal structure, the tidal dissipation is computed along the entire stellar evolution. Tidal dissipation is separated into two components: the dissipation of the equilibrium (non-wavelike) tide and the dissipation of the dynamical (wavelike) tide. For evolved stars the dynamical tide is constituted by progressive internal gravity waves. The evolution of the tidal dissipation is investigated and compared for both the equilibrium and dynamical tides.}
   {The significance of both the equilibrium and dynamical tide dissipation becomes apparent within distinct domains of the parameter space. The dissipation of the equilibrium tide is dominant when the star is large in size or the companion is far away from the star. Conversely the dissipation of the dynamical tide is important when the star is small in size or the companion is close to the star. The size and location of these domains depend on the masses of both the star and the companion, as well as the evolutionary phase.}
   {Both the equilibrium and the dynamical tides are important in evolved stars, and therefore both need to be taken into account when studying the tidal dissipation in evolved stars and the evolution of planetary or/and stellar companions orbiting them.}


   \maketitle
%

\section{Introduction}
    As more and more planets are being observed, they are also found around evolved stars. As of today, 221 planets have been found around giant stars\footnote{Calculated using the Extrasolar Planet Encyclopedia \url{https://exoplanet.eu/}, for planets with a $R_\star > 3$ R$_\odot$ star.}. These planets are found in a wide range of orbital periods, ranging from a few days to a few years \citep[e.g.][]{Sato2008, Nowak2013, Dollinger2021, Saunders2022, Lee2023, Pereira2024}, and there is even an observed planet that is situated so close to its host star that it should have been swallowed by the star in earlier stages of its evolution under normal stellar evolution scenarios \citep{Hon2023}. Not only planets around evolved stars have gotten more recent attention, also stellar companions have increasing importance. For instance, for Asymptotic Giant Branch (AGB) stars or planetary nebula, the presence of a binary companion is important in shaping the outflows of these stars \citep{Garcia-Segura2016,Decin2020}. Binary stars have also been observed around post-AGB stars, where the orbital properties of these binaries show significant eccentricity, which is not predicted by current binary evolution models \citep{VanWinckel2003,Oomen2018}. For these systems it is important to understand tidal effects during the late stages of stellar evolution, as tides play a significant role in the evolution of the orbital architecture of the system, as well as the rotational evolution of the objects in the system \citep[e.g.][]{Ogilvie2014,Bolmont2016,Mathis2018,Ahuir2021b}.

    Tidal dissipation is a complex process characterized by two components: the equilibrium and dynamical tides. The former arises from the hydrostatic displacement induced by the ellipsoidal deformation triggered by a companion. The energy associated with the equilibrium tide is dissipated through turbulent friction in convective layers, leading to the transfer of angular momentum between the stellar spin and its orbital motion \citep{Zahn1966a,Zahn1989a,Zahn1989b, Remus2012,Ogilvie2013,Barker2020}. This well-established mechanism has been studied in various contexts, using observations of Red Giant Branch (RGB) binaries \citep{Verbunt1995,Beck2018,Beck2022,Beck2024}, as well as theoretically for AGB binaries \citep{Mustill2012, Madappatt2016, Garcia-Segura2016} as short period AGB binaries are difficult to observe \citep{Decin2020}. Studies for AGB binaries still use parameterised equations using fixed input parameters rather than ab-initio dissipation calculations. The dynamical tide involves tidal dissipation due to the excitation of stellar oscillations by the tidal potential. In the stellar convective zone, these waves are excited in the form of inertial waves that are restored by the Coriolis force \citep[e.g.][]{Wu2005a,Wu2005b,Ogilvie2004,Ogilvie2007}, or in the form of internal gravity waves (IGW) restored by the buoyancy in the radiative zone \citep[e.g.][]{Zahn1975,Goldreich1989,Terquem1998,Goodman1998,Barker2010,Ahuir2021a}. This dynamical tide dissipation varies over several orders of magnitude depending on the structure and the rotation of stars all along their evolution \citep{Mathis2015,Gallet2017,Bolmont2017,Barker2020,Ahuir2021a}. While tidal effects have been studied extensively in the pre-main-sequence (PMS) and main-sequence (MS), these dependence's have yet to be systematically evaluated along the evolved phases, despite its already identified crucial role for subgiants \citep{Weinberg2017,Beck2018} and RGB stars \citep{Ahuir2021a}. Therefore our focus in this investigation is on computing the tidal dissipation strengths in stars throughout their entire lifetime, with a particular emphasis on the evolved phases.

    In this study, we bridge this gap by conducting a comprehensive investigation into the equilibrium and dynamical tide dissipation during the late stages of evolution. Utilising established theoretical frameworks and a grid of stellar evolution models with initial masses between 1 and 4 M$_\odot$ to study of a range of stellar evolutionary effects such as the convective/radiative structure during the MS and presence of the helium flash. We aim to quantify and compare the dissipation strengths of the equilibrium and dynamical tides, providing a complete understanding of their contributions to the overall dissipation of tidal energy along the star's lifetime.

    In Sect~\ref{sec:TidalDissipationModel}, we provide an overview of the type of waves that can be excited in stars by tides, as well as an overview of the theoretical frameworks used to model the dissipation of the equilibrium and dynamical tides. In Sect.~\ref{sec:StellarEvolution}, we describe the stellar evolution models used in this study and their physical ingredients. In Sect.~\ref{sec:TidalDissipationAlongEvolution}, we investigate the equilibrium and dynamical tide dissipation as well as their relative strength along the stellar evolution. Finally, in Sect.~\ref{sec:Conclusion} summarizes the conclusions.

\section{Tidal dissipation modelling}\label{sec:TidalDissipationModel}
    In this section, we provide an overview of the theoretical frameworks used to compute the equilibrium and dynamical tides and their dissipation.
    \subsection{General framework}
        Considering two bodies, for instance a star and a planet, or two stars, where only the deformation of one object is considered, the gravitational potential of the secondary object can be expressed as a multipole expansion in spherical harmonics in the reference frame attached to the primary. The tidal potential is then given by the difference between the gravitational potential induced by the secondary object at each point of the extended primary body and its value at its center of mass and the first-order term ensuring the Keplerian motion. The tidal potential can be expressed as given in \cite{Ogilvie2014}:
        \begin{equation}\setlength{\jot}{-3pt}
            \begin{aligned}
                \Psi(r, \theta, \varphi, t)=\Re \left\{\sum_{l=2}^{\infty} \sum_{m=0}^l \sum_{n=-\infty}^{\infty}\right.\frac{G M_2}{a} & A_{l, m, n} (e, i)\left(\frac{r}{a}\right)^l\\& \times Y_l^m(\theta, \varphi) \mathrm{e}^{-\mathrm{i} n \Omega_o t}\Bigg\}\ ,
            \end{aligned}
        \end{equation}
        where $M_2$ is the mass of the companion object (in g), $a$ is the semi-major axis of the orbit (in cm), $e$ is the eccentricity of the orbit, $i$ is the inclination of the orbit, $r,\theta,\varphi$ are spherical coordinates centered at the origin of the reference frame attached to the primary center of mass, $Y_l^m$ are the spherical harmonics, $\Omega_o=\sqrt{G(M_1 + M_2)/a^3}$ is the orbital frequency (in rad/s), and $A_{l, m, n}$ are the tidal coefficients (see table 1 in \citealt{Ogilvie2014}). In this study we focus on a coplanar circular orbit. In this case only the $l=m=n=2$ terms are non-zero in the quadrupolar approximation ($l>2$ is neglected; we refer the reader to \citealp{Mathis2009} for the conditions ruling this approximation), and the tidal potential can be expressed as:\vspace*{-2pt}
        \begin{equation}\label{eq:TidalPotential}
            \begin{aligned}
                \Psi=&\operatorname{Re} \left\{\frac{G M_2}{a} \sqrt{\frac{6\pi}{5}}\left(\frac{r}{a}\right)^2 Y_{2}^{2}(\theta, \varphi) \mathrm{e}^{-\mathrm{i} 2 \Omega_o t} \right\}\\
                 \equiv &\ \varphi_T(r)\operatorname{Re} \left\{ Y_{2}^{2}(\theta, \varphi) \mathrm{e}^{-\mathrm{i} 2 \Omega_o t} \right\}\ .
            \end{aligned}
        \end{equation}
        In the rest of the text, we will continue using $l$, $m$ and $n$ for clarity and consistency with the literature. Since $Y_l^m \propto e^{im\varphi}$, the tidal potential has as complex argument $m\varphi - n\Omega_o t$. This means that the tidal potential rotates with a frequency $\omega_t=n\Omega_o-m\Omega_s$, where $\omega_t$ is the tidal frequency and $\Omega_s$ is the spin frequency of the star (in rad/s), which is assumed to be uniform here. This tidal frequency will be the characteristic frequency of the tidal waves excited in the star.

        Tidal dissipation can be expressed through multiple formalisms, using for instance the tidal quality factor $Q$ \citep{Kaula1962}, the modified tidal quality factor $Q^\prime$ \citep{Ogilvie2007} or the Love number $k_l^m$ \citep{Love1911}. The tidal quality factor is defined as the ratio between the energy stored in the tidal bulge and the energy dissipated per orbit. The Love number is the ratio between the perturbation of the primary's gravitational potential induced by the presence of the companion and the tidal potential, evaluated at the stellar surface. The tidal quality factor and the Love number are related by
        \begin{equation}
            Q_l^m(\omega_t)^{-1}=\operatorname{sgn}(\omega_t)\Im\left(k_l^m(\omega_t)\right)/\left|k_l^m(\omega_t)\right|
        \end{equation}
        and the modified tidal quality factor is related to the tidal quality factor and the Love number by
        \begin{equation}
            \frac{3}{2Q_l^{\prime m}(\omega_t)} = \frac{\Re\left(k_l^m\right)}{Q_l^m}\ ,
        \end{equation}
        where in the case of weakly dissipative stellar fluid $\left| k_{l}^{m}\right|\approx \Re\left(k_{l}^{m}\right)$. Higher values of the imaginary part of the Love number indicate stronger tidal dissipation, while lower values of the tidal quality factor indicate stronger dissipation.

    \subsection{Tidal wave excitation}
        In order to understand the tidal dissipation in the primary star, it is important to understand what types of waves can be excited in the star, and more importantly what types of waves can be excited by the tidal potential. The tidal potential is a periodic potential, and therefore it can only excite waves with a frequency equal to the tidal frequency $\omega_t$, the frequency at which the tidal potential rotates. If it is possible to excite waves at tidal frequencies, the tidal potential then triggers the so-called dynamical tide (e.g. \citealp{Zahn1975} and \citealp{Ogilvie2004}) and the related dissipation and angular momentum exchanges. In this section the different types of waves that can be excited by tides in a primary evolved low or intermediate-mass evolved star are discussed, assuming a companion on a circular coplanar orbit at a distance of 1 AU.

        \subsubsection{Inertial waves}
            Inertial waves are waves that have the Coriolis force as a restoring force. These waves can only be excited in rotating stars for sufficiently low frequencies in the regime $\omega \in [-2\Omega_s, 2\Omega_s]$ \citep{Rieutord2015}. For tides to excite inertial waves this criterion holds if $\Omega_o < 2\Omega_s$. This means that for sufficiently slowly rotating stars, the tidal potential cannot excite inertial waves.
            
            Stars with a convective envelope during the Main Sequence (MS) phase are spun down because of the magnetic breaking by their pressure-driven winds \citep[e.g.][]{Skumanich1972,Kawaler1988}. Therefore, if there is no sufficiently massive companion to spin up the star, the tidal potential cannot excite inertial waves during their subsequent evolution since inertial waves become less important already when progressing along the main-sequence \citep{Mathis2015,Gallet2017}. Stars without a convective envelope during the MS (stars with an initial mass above $\approx$ 1.4 M$_\odot$) don't experience this spin down, and are expected to still have significant rotation rates during their subsequent evolution. However, observations of these stars reveal low rotation rates during the RGB, the cause of which thought to be stronger rotational damping than predicted by models, or differential rotation within the star \citep{Ceillier2017}. Rotation during the AGB phase has been investigated theoretically but includes only the equilibrium tide. Rotation rates found in these studies predict sufficient spin up (but still low rotation rates) so that inertial waves can be excited for stellar mass companions \citep{Garcia-Segura2016,Madappatt2016}, but not in the case of planetary companions \citep{Madappatt2016}.

            Depending on the internal structure of the star, different types of inertial waves can be excited. When the star is fully convective, global normal modes can be excited \citep{Wu2005a,Wu2005b}, while in stars with a radiative interior, wave attractors can be excited due to the excitation and reflection of the waves at the boundaries of the radiative zone (at the so-called critical latitude; \citealp{Ogilvie2013}). The strength of the dissipation of these waves depends on the Ekman number $\mathrm{Ek}=\nu_t/(2\Omega_s R_\star^2)$, with $\nu_t$ the turbulent viscosity (in cm$^2$/s) and $R_\star$ the radius of the primary star (in cm). For low Ekman numbers, the dissipation of these waves is strong, while for high Ekman numbers, the resonant dissipation of these waves is suppressed \citep{Ogilvie2004,AuclairDesrotour2015}. For evolved stars, the Ekman number is typically on the order of $10^{-3} \rightarrow 10^{0}$ (see App. \ref{sec:Ekman}). This is sufficiently high to damp the resonant dissipation of these waves.

            Tidal dissipation from inertial waves can also be calculated using its frequency-averaged value \citep{Ogilvie2013,Mathis2015}. In this way, the tidal dissipation of inertial waves can be evaluated analytically when assuming bi-layers stellar models with averaged density for each zone. The dissipation is dependent on the square of the rotation frequency normalised by the break up rotation frequency as well as on the fifth power of the radial aspect ratio defined as the radius of the radiative core divided by the stellar radius ($R_c/R_\star$). As evolved stars are slow rotators, as well as they have a small core compared to the stellar radius (see Sect. \ref{sec:InternalStructure}), the dissipation from inertial waves are negligible.

            As the tidal frequency is not always in the inertial wave regime, and in the case it is the dissipation of these waves are expected to be low for evolved stars, the dissipation of inertial waves is not considered in this study. As evolved stars have slow rotation rates for single stars, while their rotation is dependent on the mass of the secondary in binaries because of potential tidal spin up \citep{Garcia-Segura2016,Madappatt2016}, the primary star will be considered as non-rotating for simplicity in the rest of the study.

        \subsubsection{Pressure waves}
            \begin{figure}
                \centering
                \includegraphics[width=\linewidth]{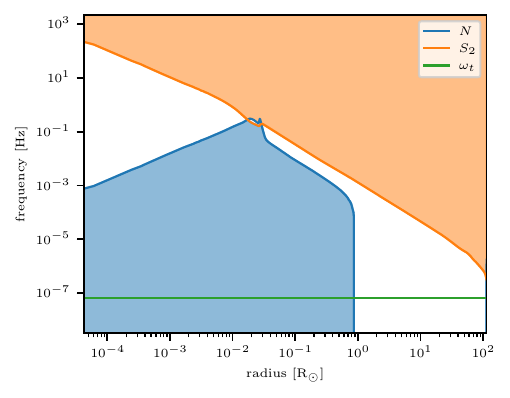}
                \caption{The Brunt-Väisälä ($N$, in blue), Lamb (for $l=2$; $S_2$ in orange) and tidal ($\omega_t$ in green) frequencies for a planetary companion orbiting an RGB star with $M_\text{ZAMS} = 1$ M$_\odot$ at a distance of 1 AU. The coloured regions indicate the regions where the different types of waves can be excited (see text for details).}
                \label{fig:N2}
            \end{figure}
            Pressure waves (p-waves) are waves that have pressure as their restoring force. These waves can be excited at frequencies higher than the Lamb-frequency $S_l = \frac{l(l+1)c_s^2}{r^2}$ (in Hz), with $c_s$ being the local sound speed \citep[][in cm/s]{Aerts2010}. The Lamb-frequency variation for $l=2$ as a function of radius is shown in Fig.~\ref{fig:N2} for a star with initial mass of 1 M$_\odot$ during the RGB phase. Here the tidal frequency is also shown for a planet on a circular orbit at a distance of 1 AU. As can be seen, the tidal frequency is always lower than the Lamb-frequency ($\omega_t < S_l$), and therefore no pressure waves can be excited by the tidal potential. This conclusion holds for all stars within our considered mass range, and therefore the dissipation of pressure waves is not considered in this study.
        
        \subsubsection{Gravity waves}
            Gravity waves (g-waves) are waves that have buoyancy as their restoring force. These waves can be excited at frequencies lower than the Brunt-Väisälä frequency $N$ \citep[in Hz;][]{Aerts2010}:
            \begin{equation}
                N^2=g_0\left(\frac{\partial_r p_0}{\Gamma_1 p_0}-\frac{\partial_r \rho_0}{\rho_0}\right)\ ,
            \end{equation}
            where $g_0$, $p_0$ and $\rho_0$ are the unperturbed gravitational acceleration (in cm/s$^2$), pressure (in g/cm s$^2$), and density (in g/cm$^3$) respectively. $\Gamma_1 = (\partial \ln p_0 / \partial \ln \rho_0)_S$ is the first adiabatic exponent, where $S$ is the macroscopic entropy.
            This Brunt-Väisälä frequency is shown as a function of radius in Fig.~\ref{fig:N2} for a star with initial mass of 1 M$_\odot$ during the AGB phase. Here the tidal frequency is also shown for a planet on a circular orbit at a distance of 1 AU. As can be seen, the tidal frequency is much lower than the maximal Brunt-Väisälä frequency ($\omega_t \ll N_\text{max}$), and therefore gravity waves in the form of internal gravity waves (IGW) can be excited by the tidal potential, and need to be taken into account when computing the dynamical tide and its dissipation.

            Depending on the efficiency of the radiative damping, the waves are either damped before they can be reflected at the boundaries of the radiative zone, or they are reflected at the boundaries and travel back and forth through the radiative zone. The former type of waves are called progressive gravity waves, while in the latter case, the reflecting interaction of the waves create standing gravity modes. The two cases are separated by the critical frequency $\nu_\mathrm{c}$, which is the frequency at which the radiative damping is strong enough to damp the waves with a factor $e$ before they can be reflected at the boundaries of the radiative zone. The critical frequency (in Hz) can be expressed as \citep{Alvan2015}:\vspace*{-4pt}
            \begin{equation}\label{eq:CriticalFrequency}
                \nu_\mathrm{c} = [l(l+1)]^{\frac{3}{8}}\left(\left| \int_{r_\text{in}}^{r_\text{out}} K_{\mathrm{T}} \frac{N^3}{r_1^3} \mathrm{~d} r_1\right|\right)^{\frac{1}{4}}\ ,
            \end{equation}
            where $K_{\mathrm{T}}$ is the thermal diffusivity (in cm$^2$/s, calculated following \citealp{Viallet2015}). Frequencies lower than the critical frequency are progressive gravity waves, while frequencies higher than the critical frequency are gravity modes.
            \begin{figure}
                \centering
                \includegraphics[width=\linewidth]{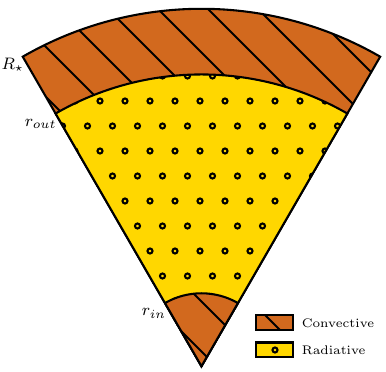}
                \caption{Schematic representation of the radiative and convective shells in the case of the three-layer structure used in this study. Radii not up to scale for a real stellar structure (See Kippenhahn diagrams e.g. Fig.~\ref{fig:Kippenhahn}).\vspace*{-4pt}}\label{fig:TrilayerStructure}
            \end{figure}
            Here $r_\text{in}$ and $r_\text{out}$ are the inner and outer boundaries of the radiative zone assuming a three-layer structure (see Fig.~\ref{fig:TrilayerStructure}). In the case there is only a radiative core and a convective envelope, $r_\text{in} = 0$ and $r_\text{out} = r_\text{{c}}$. In the case there is a convective core and a radiative envelope, $r_\text{in} = r_\text{{c}}$ and $r_\text{out} = R_{\star}$. In the case of a three-layer structure (such as in MS F-type stars, or horizontal branch stars), one needs to be careful with this definition as waves starting at the inner boundary may interact with waves starting at the outer boundary, and still create g-modes while being in the progressive wave regime.

            When the amplitude of the gravity waves become sufficiently large, the waves can start to feel non-linear effects known as wave breaking \citep{Barker2010,Barker2020}. In this case an absorption barrier is created, such that the waves cannot be reflected, and the waves are absorbed. When this occurs the waves are not able to form standing modes, and our formalism still holds for frequency above the critical frequency.

            The critical frequency in evolved stars is higher than the tidal frequency (see Sect.~\ref{sec:Pcrit}), and therefore the tidal potential will mostly excite progressive gravity waves in evolved stars.

    \subsection{Tidal dissipation}
        \subsubsection{Equilibrium tide}\label{sec:Equilibrium}
            The equilibrium tide is the tidal dissipation originating from the hydrostatic deformation of an object due to the gravitational potential of a companion \citep{Zahn1966a,Zahn1989a}. The equilibrium tide is dissipated through turbulent friction in convective layers. In order to calculate the dissipation of the equilibrium tide, the tidal displacement of the star needs to be calculated. To calculate this displacement, the non-wavelike component of the gravitational potential $\Phi_l^\text{nw}$ needs to be obtained. This is done by solving the differential equation \citep[e.g.][]{Zahn1966a,Dhouib2024}:
            \begin{equation}
                \frac{1}{r^2}\frac{\dd}{\dd r}\left(r^2 \frac{\dd \Phi_l^\text{nw}}{\dd r}\right) - \frac{l(l+1)}{r^2}\Phi_l^\text{nw} - 4\pi G \frac{\dd \rho_0}{\dd r} \frac{1}{g_0} \left(\Phi_l^\text{nw} + \Psi_l\right) = 0
            \end{equation}
            for $l=2$, where $\Psi_l$ the tidal potential (in erg). Boundary conditions are chosen to ensure regularity at the center, and continuity at the surface \citep{Ogilvie2013,Dhouib2024}:
            \begin{equation}
                \begin{cases}
                    \frac{\dd \ln \Phi_l^\text{nw}}{\dd \ln r} = l & \text{at }r=\eta R_\star\text{ for }\eta\to0\\
                    \frac{\dd \ln \Phi_l^\text{nw}}{\dd \ln r} = -(l+1) & \text{at }r= R_\star
                \end{cases}\ .
            \end{equation}
            When the non-wavelike component of the gravitational potential is known, the tidal displacement (in cm, radial $r$ and horizontal $h$ components) can be calculated following \citep{Zahn1966a,Remus2012,Dhouib2024} to ensure the equilibrium tide continuity at each radiative/convective boundary:\vspace*{-4pt}
            \begin{equation}\label{eq:Displacement}
                \xi_{r, l}^\text{nw} = -\frac{\Phi_l^\text{nw} + \Psi_l}{g_0}\ ,\ \ \ \xi_{h, l} = \frac{1}{l(l+1)}\left(2\xi_{r, l}^\text{nw} + r \frac{\dd \xi_{r, l}^\text{nw}}{\dd r}\right)\ .
            \end{equation}
            Then using these expressions we compute the dissipation of the equilibrium tide following \cite{Barker2020}:\vspace*{-4pt}
            \begin{equation}\label{eq:EquilibriumTide}
                \Im \left(k^2_2\right)_\text{eq}= \frac{16\pi G\omega_t}{4(2l+1)R_\star|\varphi_T(R_\star)|^2}\int_0^{R_\star} r^2 \rho \nu_t D_l(r) \dd r\ ,
            \end{equation}
            with\vspace*{-4pt}
            \begin{equation}\label{eq:D_l}\setlength{\jot}{-2pt}
                \begin{aligned}
                    D_l(r) = \frac{1}{3} & \left(3\frac{\dd \xi_{r, l}^\text{nw}}{\dd r} - \frac{1}{r^2}\frac{\dd(r^2 \xi_{r, l}^\text{nw})}{\dd r} + l(l+1)\frac{\xi_{h, l}^\text{nw}}{r}\right)^2 \\
                    & + l(l+1)\left(\frac{\xi_{r, l}^\text{nw}}{r} + r \frac{\dd (\xi_{h, l}^\text{nw}/r)}{\dd r}\right)^2 \\
                    & + (l-1)l(l+1)(l+2)\left(\frac{\xi_{h, l}^\text{nw}}{r}\right)^2\ ,
                \end{aligned}
            \end{equation}
            where $l=2$ and $\nu_t(x)$ the turbulent viscosity (in cm$^2$/s) given by \cite{Duguid2020}:\vspace*{-4pt}
            \begin{equation}\label{eq:TurbulentViscosity}
                \nu_t=V_{\mathrm{c}} l_{\mathrm{c}} F(\omega_t),\ 
                F(\omega_t)=\begin{cases}
                5 & |\omega_t|t_{\mathrm{c}}<10^{-2} \\
                \frac{1}{2}\left(|\omega_t| t_{\mathrm{c}}\right)^{-\frac{1}{2}} & |\omega_t| t_{\mathrm{c}} \in\left[10^{-2}, 5\right] \\
                \frac{25}{\sqrt{20}}\left(|\omega_t| t_{\mathrm{c}}\right)^{-2} & |\omega_t| t_{\mathrm{c}}>5
                \end{cases}
            \end{equation}
            with $V_{\mathrm{c}}$ the convective velocity (in cm/s), $l_{\mathrm{c}}$ the mixing length (in cm), and $t_{\mathrm{c}}$ the convective turnover time (in s).
        
        \subsubsection{Dynamical tide for progressive IGW}
            Additionally to the equilibrium tide is the dynamical tide, constituted here by progressive internal gravity waves. These waves are excited with the frequency being the tidal frequency $\omega_t$, and are dissipated through radiative damping \citep[e.g.][]{Zahn1975,Goldreich1989}.
            
            Assuming the three-layer structure (see Fig.~\ref{fig:TrilayerStructure}), the tidal dissipation can be calculated for waves emerging from the convective core and the convective envelope (the derivation can be found in App. \ref{sec:imk2dyn}, which in the limit of a low companion mass, i.e. a planet, is equal to the formula derived by \citealp{Ahuir2021a}):
            \begin{equation}
                \begin{aligned}
                \Im \left(k^2_2\right)_\text{IGW}= & \frac{3^{-\frac{1}{3}} \Gamma^2\left(\frac{1}{3}\right)}{2 \pi} [l(l+1)]^{-\frac{4}{3}} \omega_t^{\frac{8}{3}} \frac{a^6}{GM_2^2 R_{\star}^5} \\
                    & \begin{aligned}\times\Bigg(
                        \rho_0\left(r_\text{in}\right) r_\text{in}&\left|\frac{\dd N^2}{\mathrm{~d} \ln r}\right|_{r_\text{in}}^{-\frac{1}{3}} \mathcal{F}_\text{{in}}^2\\
                        &+\rho_0\left(r_\text{out}\right) r_\text{out}\left|\frac{\dd N^2}{\mathrm{~d} \ln r}\right|_{r_\text{out}}^{-\frac{1}{3}} \mathcal{F}_\text{{out}}^2
                    \Bigg)\end{aligned}
                \end{aligned}\label{eq:IGW}
            \end{equation}
            with $\mathcal{F}_\text{{out}}$ and $\mathcal{F}_\text{{in}}$ being the tidal forcing (in cm$^2$) at the inner and outer boundary of the radiative zone, and $\Gamma$ the gamma function \citep{Abramowitz1972}. The tidal forcing term at the inner and outer boundary of the radiative zone can be expressed as \citep{Ahuir2021a}:
            \begin{equation}\label{eq:TidalForcing}
                \begin{aligned}
                \mathcal{F}_\text{{in}}&=\int_0^{r_\text{in}}\left[\left(\frac{r^2 \varphi_T}{g_0}\right)^{\prime \prime}-\frac{l(l+1)}{r^2}\left(\frac{r^2 \varphi_T}{g_0}\right)\right] \frac{X_{1, \text {in}}}{X_{1, \text {in}}\left(r_\text{{in}}\right)} \dd r \\
                \mathcal{F}_\text{{out}}&=\int_{r_\text{{out}}}^{R_{\star}}\left[\left(\frac{r^2 \varphi_T}{g_0}\right)^{\prime \prime}-\frac{l(l+1)}{r^2}\left(\frac{r^2 \varphi_T}{g_0}\right)\right] \frac{X_{1, \text {out}}}{X_{1, \text {out}}\left(r_\text{{out}}\right)} \dd r
                \end{aligned}\ ,
            \end{equation}
            where $X_{1, \text {out}}$ and $X_{1, \text {in}}$ are representations for the radial displacement originating from the inner and outer boundary of the radiative zone, and $\varphi_T$ is defined in Eq.~\eqref{eq:TidalPotential}. The radial displacement can be calculated using the following differential equations and boundary conditions \citep{Ahuir2021a}:
            \begin{equation}
                \begin{aligned}
                    &\left\{\begin{aligned}
                        & X_{1, \text {out}}^{\prime \prime}-\frac{\partial_r \rho_0}{\rho_0} X_{1, \text {out}}^{\prime}-\frac{l(l+1)}{r^2} X_{1, \text {out}}=0 \\
                        & X_{1, \text {out}}(r)_{r\to0} \propto r^{1/2+\sqrt{1/4+l(l+1)}} \\
                        & X_{1, \text {out}}^{\prime}(r)_{r\to0} \propto \left(1/2+\sqrt{1/4+l(l+1)}\right) r^{-1/2+\sqrt{1/4+l(l+1)}}
                    \end{aligned}\right.\\
                    &\left\{\begin{aligned}
                        & X_{1, \text {in}}^{\prime \prime}-\frac{\partial_r \rho_0}{\rho_0} X_{1, \text {in}}^{\prime}-\frac{l(l+1)}{r^2} X_{1, \text {in}}=0 \\
                        & X_{1, \text {out}}(r)_{r\to R_\star} \propto \rho_0\left(r-R_\star-\frac{\varphi_T(R_\star)}{g_0(R_\star)}\right) \\
                        & X_{1, \text {out}}^{\prime}(r)_{r\to R_\star} \propto \rho_0(R_\star) 
                    \end{aligned}\right.
                \end{aligned}
            \end{equation}
            where the proportionality factor in the boundary conditions is cancelled out as $X$ is always rescaled to the interaction region $X/X(r_{int})$ in Eq.~\eqref{eq:TidalForcing}. The boundary conditions are calculated by assuming that $\frac{\partial_r \rho_0}{\rho_0}$ is small close to the center of the star, and that $\frac{l(l+1)}{r^2}$ is small close to the surface of the star in the case of the studied coplanar circular orbit of the companion for which $l=m=2$.

\section{Stellar evolution models}\label{sec:StellarEvolution}
    \begin{figure}
        \centering
        \includegraphics[width=\linewidth]{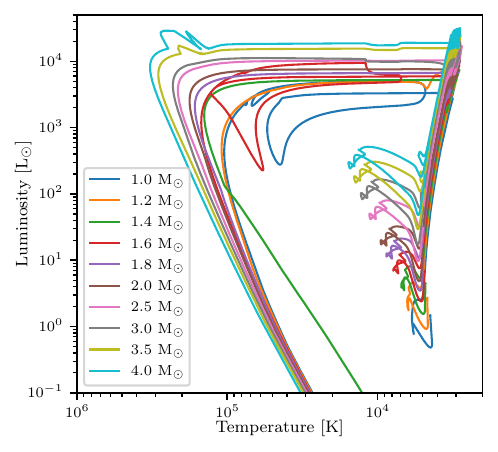}
        \vspace*{-10pt}
        \caption{Hertzsprung-Russell diagram of the stellar evolutionary models studied here. The models start in the pre-main-sequence and are evolved up to the white dwarf phase. Different colors indicate a different ZAMS mass, as indicated in the box in the left-bottom corner of the figure.\vspace*{-6pt}}\label{fig:HRD}
    \end{figure}

    To study the different tidal dissipation mechanisms throughout a star's lifetime, stellar evolutionary models are necessary. In this study we use the Modules for Experiments in Stellar Astrophysics (MESA) code \citep{Paxton2011, Paxton2013, Paxton2015, Paxton2018, Paxton2019, Jermyn2023} to calculate these stellar evolutionary models. Models are computed for Zero Age Main Sequence (ZAMS) stars whose masses vary between 1 and 4 M$_\odot$ at solar metallicity ($Z = 0.0134$; \citealp{Asplund2009}) for which the Hertzsprung-Russell (HR) diagram is shown in Fig.~\ref{fig:HRD}\footnote{The inlist used to compute the stellar evolutionary models can be found both in App.~\ref{sec:Inlist} and zenodo: \url{https://doi.org/10.5281/zenodo.11519523}}. These masses are chosen to allow the study of a range of stellar evolutionary effects such as the difference between a convective or radiative envelope during the MS and the difference between a helium flash or gradual helium burning. The models are computed from the pre-main-sequence (PMS) up to the white dwarf (WD) stage. The models are terminated when the WD is cooled down sufficiently to have a luminosity of $L = 10^{-1} $ L$_\odot$.

    To simulate convection, the Mixing Length Theory (MLT) is used following the prescription of \cite{Henyey1965}. In this prescription $\alpha_{\mathrm{MLT}}$ is the mixing length parameter, which is calibrated by \cite{Cinquegrana2022} to reproduce the solar radius and luminosity at the solar age, resulting in $\alpha_{\mathrm{MLT}} = 1.931$. As evolved stars have relative low temperatures, a dedicated low-temperature molecular opacity table \citep[\AE{}SOPUS,][]{Marigo2009} is used. For the atmosphere, a grey temperature-opacity ($T-\tau$, where $\tau$ is the optical depth) relation is assumed, based on the Eddington relation \citep{Paxton2011}.

    The mass-loss rate is calculated using the Reimers prescription \citep{Reimers1975} during the RGB phase, with a scaling factor of $\eta_\text{Reimers}=0.477$ \citep{McDonald2015}. During the AGB phase, the Bl{\"o}cker prescription \citep{Blocker1995} is used with a scaling factor of $\eta_\text{Bl{\"o}cker}=0.05$ for masses below 2 $M_\odot$ and $\eta_\text{Bl{\"o}cker}=0.1$ for masses above 2 $M_\odot$ \citep{Madappatt2016}.

    Stars undergo internal mixing, which is described by convective mixing in the convective zone. Radiative layers are also the seat of mild mixing and transport mechanisms \citep[see e.g.][]{Zahn1992,Mathis2013,Aerts2019}. To have a simple description of the mixing in the radiative zone, we assume a constant uniform mixing coefficient $D_\text{min}=10$ cm$^2$ s$^{-1}$ throughout the radiative layers. Multiple values were tested for $D_\text{min}$, and the resulting tidal dissipation was found to be insensitive to the exact value of $D_\text{min}$. This value was chosen to help numerical stability within the MESA simulations.

\section{Tidal dissipation along stellar evolution}\label{sec:TidalDissipationAlongEvolution}
    During the evolution of the star, the strength of tidal dissipation varies due to the changes in internal structure and rotation \citep[e.g.][]{Mathis2015,Gallet2017,Bolmont2017,Barker2020,Ahuir2021a}. In this section, we investigate the tidal dissipation along the stellar evolution, focusing specifically on advanced stages from the RGB to the WD stages. To verify our results, we benchmark them with the values of the dissipation of the equilibrium and dynamical tides found in \cite{Ahuir2021a} who computed of the equilibrium and dynamical tide dissipation up to the RGB phase. The comparison was done for the 1, 1.2 and 1.4 M$_\odot$ stars in App.~\ref{sec:Comparison}. We find good agreement between our results and the results of \cite{Ahuir2021a}. Before going into detail on the tidal dissipation, we validate our formalism by examining the critical period as a function of stellar mass.

    \subsection{Critical period}\label{sec:Pcrit}
        \begin{figure}
            \centering
            \includegraphics[width=\linewidth]{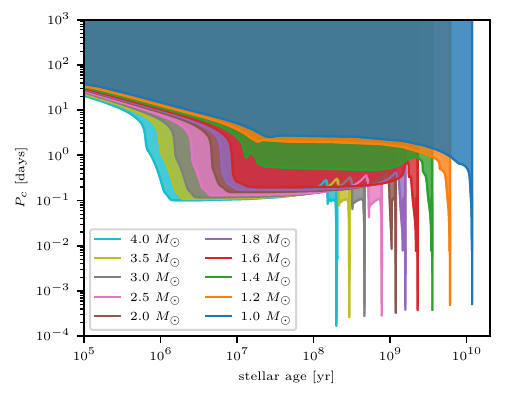}
            \caption{The critical period $P_c = 1/\nu_\mathrm{c}$ for different stellar masses. The color represents the same model as in Fig.~\ref{fig:HRD}. The coloured regions indicate the regions where the gravity waves are excited as progressive internal gravity waves.}\label{fig:Pcrit}
        \end{figure}
        The critical period ($P_c = 1/\nu_\mathrm{c}$; Eq.~\ref{eq:CriticalFrequency}) as a function of stellar age is shown in Fig.~\ref{fig:Pcrit} for all stellar evolutionary models used in this study\footnote{The critical periods in this study are lower by a factor $2\pi$ compared to \cite{Ahuir2021a} due to a unit conversion from s/rad to s.}. During the PMS, the critical period is high (higher than a few days), and decreases until the MS starts. During the MS, the critical period remains approximately constant, and decreases when the RGB phase starts. Afterwards the critical period remains low (lower then 0.1 day). This means that our formalism is valid for stars along the evolved phases.

        The critical period also depends on the initial mass of the star. For lower mass stars (e.g. the 1 M$_\odot$ model), the critical period is approximately 3 days during the MS, while for higher mass stars (e.g. the 4 M$_\odot$ model), the critical period is approximately 0.1 days during the MS. Therefore it is possible for short period companions (with periods of the order of 1 day) around the lower mass stars (below or equal to 1.2 M$_\odot$) in the MS to excite g-modes in the radiative layer instead of progressive internal gravity waves. The tidal dissipation of these g-modes is likely to be more effective than the tidal dissipation estimated in this study, since standing modes might experience enhanced dissipation efficiency, potentially amplified through resonance locking \citep[e.g.][]{Witte2002,Fuller2017}.

        In principle wave breaking can also dissipate the waves, but as the critical period is small during the evolved phases, the waves will already be damped before they can break. This is not the case for the lower mass stars in the MS phase and possibly during the sub-giant phase \citep[e.g.][]{Weinberg2017}, where the critical period is higher, and the waves can break. For an investigation of wave breaking during the MS and the sub-giant phase, we refer the reader to \cite{Barker2020} and \cite{Ahuir2021a} in their App. D.

    \subsection{Earth around the Sun}
        \subsubsection{Internal structure of a 1 M$_\odot$ star}\label{sec:InternalStructure}
            \begin{figure}
                \centering
                \includegraphics[width=\linewidth]{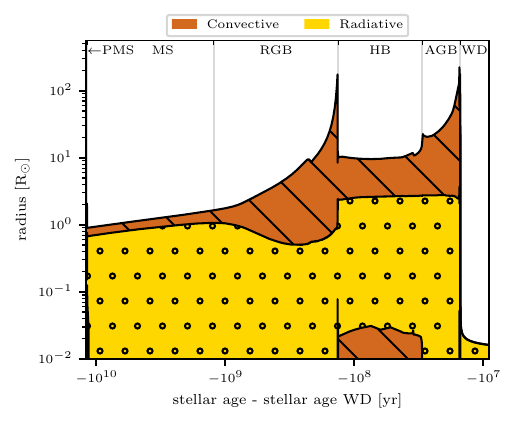}
                \caption{Kippenhahn diagram for a $M_\text{ZAMS}=1$ M$_\odot$ star. The brown hatched region represents convective layers and the yellow dotted region represents radiative layers. The upper layer represents the radius of the star. Stellar evolutionary phases (PMS to WD) are shown, and indicated in the figure.}\vspace*{-4pt}\label{fig:Kippenhahn}
            \end{figure}
            Tidal dissipation depends strongly on the internal structure of the star. Fig.~\ref{fig:Kippenhahn} shows the Kippenhahn of a $M_\text{ZAMS}=1$ M$_\odot$ star. Here the stellar age is represented as time until the end of the simulation, hence the time until the WD is cooled down to produce a luminosity of $L = 10^{-1} $ L$_\odot$. When plotting this time in logarithmic scale, the evolved phases can be shown in a single plot. The different evolutionary phases are indicated in Fig.~\ref{fig:Kippenhahn}. During the MS, the radius increases gradually, where the star has a convective envelope and a radiative core. After the MS, the star starts to expand rapidly, and the convective envelope grows. During the RGB phase, the star has a convective envelope and a radiative interior. When the star reaches the tip of the RGB, it contracts during the helium flash (brief thermal runaway nuclear burning of helium caused by the growth of the degenerate core), where the radius of the stars stays relatively constant during the horizontal branch (HB). During this phase, the star has a convective core, creating a three-layer structure. After the HB, the star expands again, and the star has a convective envelope and a radiative core in the AGB phase. When the star reaches the tip of the AGB, the star contracts again, and the star leaves the AGB track to cool down as a WD. During the WD phase, the star has a radiative interior.

        \subsubsection{Tidal dissipation}\label{sec:TidalDissipation}
            \begin{figure}
                \centering
                \includegraphics[width=\linewidth]{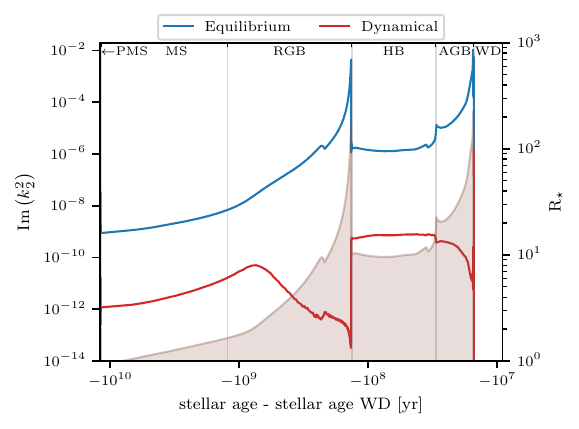}
                \caption{On the left axis the complex part of the Love number ($\Im (k^2_2)$) for both equilibrium (blue) and dynamical (red) tides is shown as function of stellar age for a $M_\text{ZAMS}=1$ M$_\odot$ star with a 1 M$_\text{Earth}$ companion at 1 AU. On the right axis the stellar radius is shown in light brown. Stellar evolutionary phases (PMS to WD) are shown, and indicated in the figure.\\}\label{fig:Imk2}
                \centering
                \includegraphics[width=\linewidth]{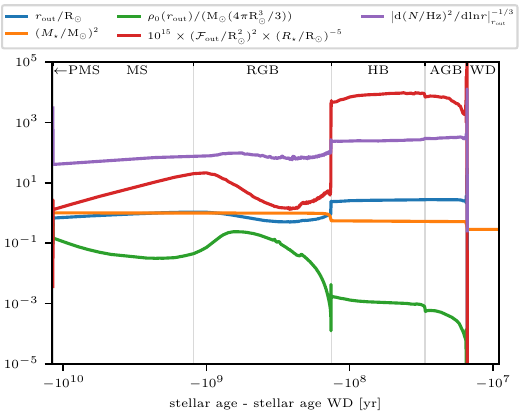}
                \caption{Different components appearing in the dissipation of the dynamical tide (Eq.~\ref{eq:IGW}) as a function of stellar age. The change in radius of the radiative-convective boundary is represented in blue, the change in stellar mass in orange, the change in local density at the boundary in green, the change in tidal forcing compared to changes in stellar radius in red, and change in the derivative of the Brunt-Väisälä frequency in purple. Stellar evolutionary phases (PMS to WD) are shown, and indicated in the figure.}\label{fig:Components}
            \end{figure}
            We consider a planet with an orbit of 1 yr and the mass of the Earth (e.g. the Earth), which is rotating around such a star with an initial mass of 1 M$_\odot$ (e.g. the Sun). For simplicity the orbit is assumed to remain at a period of 1 yr. In Fig.~\ref{fig:Imk2}, we compute the evolution of the imaginary part of the Love number both for the equilibrium and dynamical tides as a function of time. Looking at the equilibrium tide in the figure, the Love number follows the trend of the radius quite well. When the radius increases, the Love number increases, and vice versa. This is due to the fact that for stars with a larger radius, the local gravity is weaker, and therefore the star is more easily deformed by the tidal potential. This results in larger values of $\xi_r$ and $\xi_h$, and a larger integral in Eq.~\eqref{eq:EquilibriumTide}. This is compensated by the dependance of $R_\star^{-(2l+1)}$ (including the $R_\star^l$ dependence of $\varphi_T(R_\star)$) from the denominator in the equation. Inside the integral the dependance on $R_\star$ is a bit more complicated. Starting with $\xi_r$ and $\xi_h$, which are dependant on $R_\star^4$ ($R_\star^2$ from the tidal potential, and $R_\star^2$ from the reduction of the local gravity; see Eq.~\ref{eq:Displacement}), which appears squared divided by the radius inside $D_l$ (see Eq.~\ref{eq:D_l}). This results in a $R_\star^6$ dependance of $D_l$, and a $R_\star^8$ dependance inside the integral (from the additional $r^2$ in the integral). In total the integral is thus dependant on $R_\star^9$ (due to the integration), resulting in a $R_\star^4$ dependance of the equilibrium tide dissipation (in agreement with \citealp{Remus2012}).
            
            This is not the case for the dynamical tide. To explain the complex behaviour of its dissipation, the different components appearing in the dissipation of the dynamical tide (Eq.~\ref{eq:IGW}) are shown as a function of stellar age in Fig.~\ref{fig:Components}. The change in radius of the radiative-convective boundary is represented in blue, the change in stellar mass in orange, the change in local density at the radiative-convective boundary in green, the change in tidal forcing compared to changes in stellar radius in red, and change in the derivative of the Brunt-Väisälä frequency in purple. The change in radius of the radiative-convective boundary, change in stellar mass and change in the derivative of the Brunt-Väisälä frequency are small compared to the changes in the tidal forcing and local density at the boundary. This is due to the low power of the Brunt-Väisälä frequency and the boundary radius, as well as the small change in the stellar mass. The dissipation is proportional to $R_\star^{-5}$, and thus the dynamical tide dissipation responds inversely to an increase in radius\footnote{Note that although the Love number is dependent on $R_\star^{-5}$, when calculating the change of the semi-major axis of the orbit throughout time there is an additional factor of $R_\star^{5}$ cancelling out this direct dependence on stellar radius.}. This effect is compensated by the tidal forcing $\mathcal{F}$ which depends on $\varphi_T$ that scales as $R_\star^{2}$. When multiplying their contributions, a complex pattern appears (as can be seen in red in Fig.~\ref{fig:Components}). First this multiplication coefficient ($\mathcal{F}^2/R_\star^5$) increases, as the star increases while the size of the convective envelope remains approximately constant. When the convective envelope starts to becomes thicker at the start of the RGB, the tidal forcing grows slower than the stellar radius increases, as $X(r)/X_\mathrm{out}$ decreases as $X_\mathrm{out}$ increases for larger sizes of the convective envelope, and therefore the multiplication coefficient decreases. At the same time the local density at the boundary layer decreases as well, due to the large extend (and thus low density) of the envelope. This combination results in a dissipation of the dynamical tide that first increase during the subgiant phase and at the start of the RGB (in agreement with \citealp{Ahuir2021a}) and then decrease during the RGB (see Fig.~\ref{fig:Imk2}). When the helium flash occurs and the star contracts, the extent of the radiative interior increases again, and therefore the tidal forcing increases. For that reason, the dynamical tide dissipation is stronger during the HB compared to the dissipation during the RGB. Additionally, the star has a small convective core in this phase, and there is a contribution from the inner boundary of the radiative zone to the tidal dissipation, but this component is negligible compared to the contribution from the outer boundary. During the HB, the internal structure remains approximately constant, and therefore the tidal dissipation remains approximately constant. When the star starts to expand during the AGB, the size of the radiative core remains approximately constant, and therefore the multiplication coefficient remains approximately constant as well. However, as the stellar radius grows, and therefore the density at the boundary layer decreases, the dynamical tide dissipation decreases in strength. During the WD phase, the star is fully radiative. In this case tidally forced (travelling) waves can still be excited, but not at a radiative-convective boundary, and thus our formalism does not apply anymore. The dynamical tide can now excite standing gravity modes \citep{Fuller2011,Fuller2012,Fuller2013,Fuller2014} for stellar companions or f-modes \citep{Veras2019} for planetary companions. These dynamical tides are not calculated in this study.
            
            Overall the tidal dissipation of the equilibrium tide is dominant compared to the dissipation of the dynamical tide at this orbital period. But as they have different dependencies this will change for different orbital periods and different primary and secondary masses, which is shown in the next sections.

        \subsection{Dependence of tidal dissipation on orbital period}
            The strength of the equilibrium and dynamical tides are dependent on the orbital distance. For the equilibrium tide dissipation this is captured in the linear dependence on $\omega_t$, but there is also a complex dependence of the turbulent viscosity on $\omega_t$ (see Eq.~\ref{eq:TurbulentViscosity}). For the dynamical tide dissipation there is a dependence in $\omega_t$, $\Omega_o$, the tidal forcing, and the boundary conditions of $X$. This results in a complex dependence on orbital period which is investigated for a 1 M$_\text{Jup}$ planet around a 1 M$_\odot$ star.

            \subsubsection{Equilibrium tide}\label{sec:EquilibriumTide}
                \begin{figure}
                    \centering
                    \includegraphics[width=\linewidth]{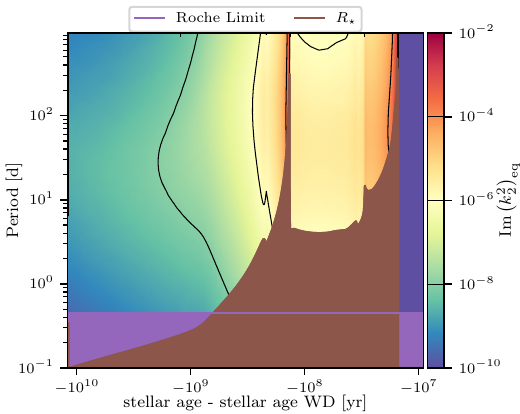}
                    \caption{Complex part of the Love number for the equilibrium tide ($\Im (k^2_2)_\text{eq}$) as a function of the orbital period and stellar age, for a $M_\text{ZAMS}=1$ M$_\odot$ star. $R_\star$ (in brown) represents the orbital period for which a planet orbits at the surface of the star, and the Roche limit is given in purple. Changes in stellar evolutionary phase are indicated with a ticks on the upper axis.}\label{fig:Imk2Eq}
                \end{figure}
                \begin{figure}
                    \centering
                    \includegraphics[width=\linewidth]{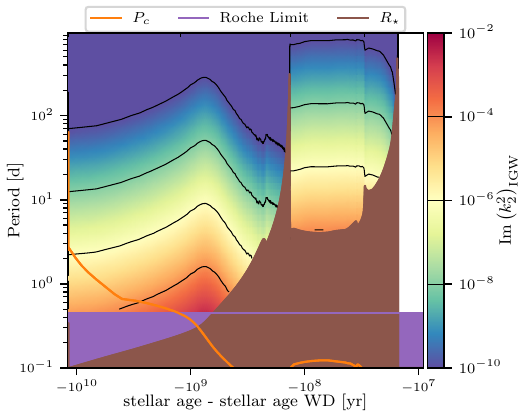}
                    \caption{Complex part of the Love number for the dynamical tide ($\Im (k^2_2)_\text{IGW}$) as a function of orbital period and stellar age, for a $M_\text{ZAMS}=1$ M$_\odot$ star. The critical period $P_c$ is shown in orange, for which planets orbiting above this period excite progressive internal gravity waves. $R_\star$ (in brown) represents the orbital period for which a planet orbits at the surface of the star, and the Roche limit is given in purple. Changes in stellar evolutionary phase are indicated with a ticks on the upper axis.}\label{fig:Imk2IGW}
                \end{figure}
                The dissipation of the equilibrium tide as a function of stellar age and orbital period is shown in Fig.~\ref{fig:Imk2Eq}. When a companion is too close to the star, the companion will be destroyed by tidal forces. This is represented by the Roche limit, which is shown for a $M=1$ M$_\text{Jup}$, $R=1$ R$_\text{Jup}$ planet in Fig.~\ref{fig:Imk2Eq} in purple (for the details of its calculation, we refer to \citealp{Benbakoura2019}). When the Roche limit is located inside the star, it is no longer a relevant parameter. In this case a planet might plunge inside the star and reach deep layers before being completely destroyed \citep{Lau2022}. Here it can be seen that by increasing the orbital period, the equilibrium tide dissipation increases until a maximum is reached, after which the equilibrium tide dissipation becomes weaker again. The reason is that for low orbital periods, the tidal frequency is sufficiently high such that in the dominant region (the region where most of the dissipation occurs) the turbulent viscosity is proportional to the inverse square of the tidal frequency. Because there is an additional linear dependence of the tidal frequency on the Love number of the equilibrium tide, the dissipation of the equilibrium tide is proportional to the inverse of the tidal frequency, and hence will increase with increasing orbital period. However, when the orbital period is sufficiently high, the tidal frequency times the convective time $t_c$ becomes high enough (see Eq.~\ref{eq:TurbulentViscosity}) such that the turbulent viscosity is proportional to the inverse square root of the tidal frequency, and thus the dissipation of the equilibrium tide is proportional to the square root of the tidal frequency. Therefore, by increasing the orbital period further, the equilibrium tide dissipation will decrease. The final regime of Eq.~\eqref{eq:TurbulentViscosity}, where the turbulent viscosity is independent of the tidal frequency, is not reached in this study, as the orbital periods are not high enough.

            \subsubsection{Dynamical tide}
                The dynamical tide dissipation as a function of stellar age and orbital period is shown in Fig.~\ref{fig:Imk2IGW}. Here it can be seen that by increasing the orbital period, the dynamical tide dissipation strictly decreases. Looking at Eq.~\eqref{eq:IGW}, there is a dependence on $\omega_t$, $\Omega_o$, the tidal forcing, and the boundary conditions of $X$. The tidal forcing is proportional to the the square of the orbital period (as there is $\varphi_T$, which has a dependence of $a^{-3}\propto \Omega_o^{2}$). Hence, this cancels out the direct dependence on the orbital frequency in the equation. The dependence in the boundary condition of $X$ on the orbital period (again due to $\varphi_T$) remains small, and only has a small effect at low orbital periods. The dependence on $\omega_t^{8/3}$ is the dominant factor, and thus the dynamical tide dissipation is proportional to $P_\text{orb}^{-8/3}$. At the end of the star's evolution, when the star is a WD and the star is fully radiative, the dynamical tide is not calculated as our formalism does not hold. Therefore this region is left blank in Fig.~\ref{fig:Imk2IGW}. In this region, the dynamical tide can be composed of standing g-modes and f-modes \citep{Fuller2011,Fuller2012,Fuller2013,Fuller2014,Veras2019} and thus its dissipation will have a different dependence on the orbital period.\vspace*{-15pt}
                \begin{figure}
                    \centering
                    \includegraphics[width=\linewidth]{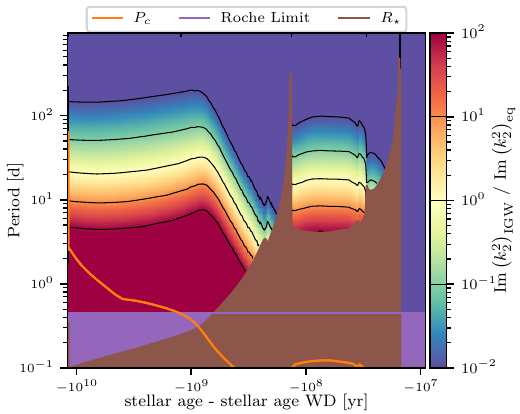}
                    \caption{Ratio of the complex parts of the Love number for the dynamical to equilibrium tide ($\Im (k^2_2)_\text{IGW}/\Im (k^2_2)_\text{eq}$) as a function of orbital period and stellar age, for a $M_\text{ZAMS}=1$ M$_\odot$ star. The critical period $P_c$ is shown in orange, for which planets orbiting above this period excite progressive internal gravity waves. $R_\star$ (in brown) represents the orbital period for which a planet orbits at the surface of the star, and the Roche limit is given in purple. Changes in stellar evolutionary phase are indicated with a ticks on the upper axis.\vspace*{-10pt}}\label{fig:Imk2Ratio}
                \end{figure}

            \subsubsection{Relative strengths}
                Since the strengths of the equilibrium and dynamical tide dissipation have a different dependence on stellar age and orbital period, the dissipation of the equilibrium tide will be dominant in some regions of the parameter space, while in other regions, the dissipation of the dynamical tide will be more important. This can be seen in Fig.~\ref{fig:Imk2Ratio}, where their ratio is shown. During the MS, the dynamical tide dissipation is dominant (a factor 10 higher than the equilibrium tide dissipation) for orbits shorter than a 10 days in agreement with previous work \citep{Ahuir2021a}. There is a gradual change until orbital periods longer than 50 days where the equilibrium tide dissipation dominates. When the star enters the RGB, the equilibrium tide dissipation increases, while the dynamical tide dissipation decreases (see Sect.~\ref{sec:TidalDissipation}), resulting in the equilibrium tide dissipation becoming dominant for shorter orbital periods, and dominating completely when the star is sufficiently large. When the star contracts and becomes a HB star, the dynamical tide dissipation becomes stronger, remaining relevant again for orbits up to 30 days. When the star enters the AGB, the equilibrium tide dissipation becomes dominant again, similar to the RGB phase. When the star contracts to become a WD, the star becomes fully radiative and our formalism cannot be applied anymore. Then, the dissipation of the equilibrium tide dominates. However, in such a configuration tidal g- and f- modes can be excited and their dissipation may dominate the one of the equilibrium tide.

        \subsection{Dependence of tidal dissipation on secondary mass}
        \begin{figure*} 
            \centering
            \includegraphics[width=\linewidth]{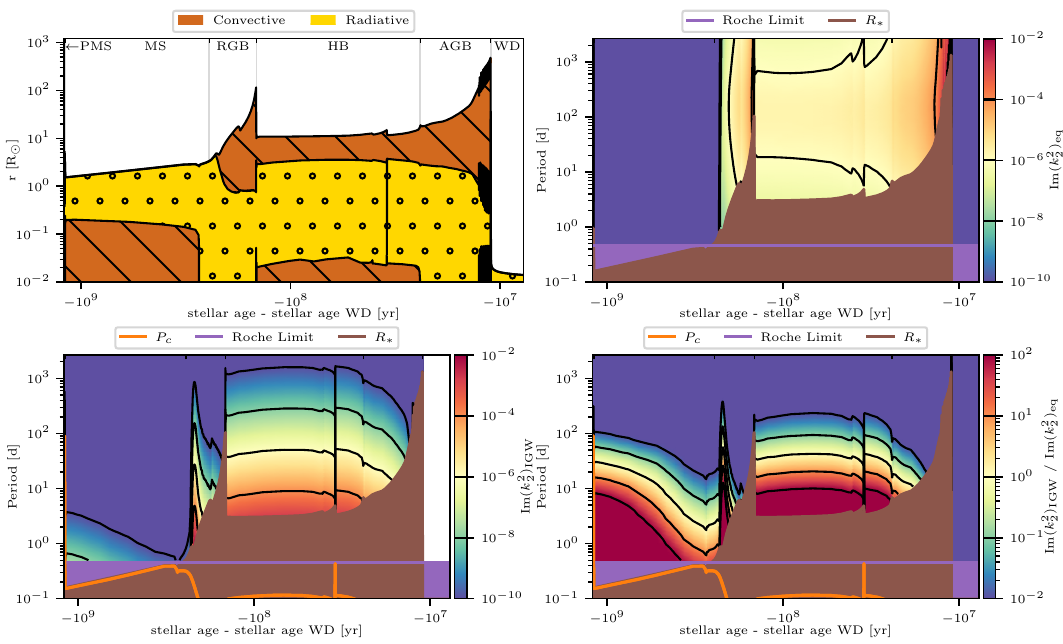}
            \caption{Figures for a $M_\text{ZAMS}=2$ M$_\odot$ star. On the top left a Kippenhahn diagram is shown, where the brown hatched regions represents convective layers and the yellow dotted region represents radiative layers. Stellar evolutionary phases (PMS to WD) are shown, and indicated in the figure. On the top right, bottom left and bottom right the complex part of the Love number for the equilibrium tide ($\Im (k^2_2)_\text{eq}$), the dynamical tide ($\Im (k^2_2)_\text{IGW}$) and the ratio of the dissipation's of the dynamical tide to the dissipation of the equilibrium tide ($\Im (k^2_2)_\text{IGW}/\Im (k^2_2)_\text{eq}$) are shown, respectively, as a function of stellar age and orbital period. Here the Roche limit is shown in purple, the critical period ($1/\nu_\mathrm{c}$, see Eq.~\ref{eq:CriticalFrequency}) in orange, and the period at which a planet orbits at the stellar radius in brown. Here changes in stellar evolutionary phase are indicated with a ticks on the upper axis.\vspace*{-5pt}}\label{fig:FullM20}
        \end{figure*}
        \begin{figure*}
            \centering
            \includegraphics[width=\linewidth]{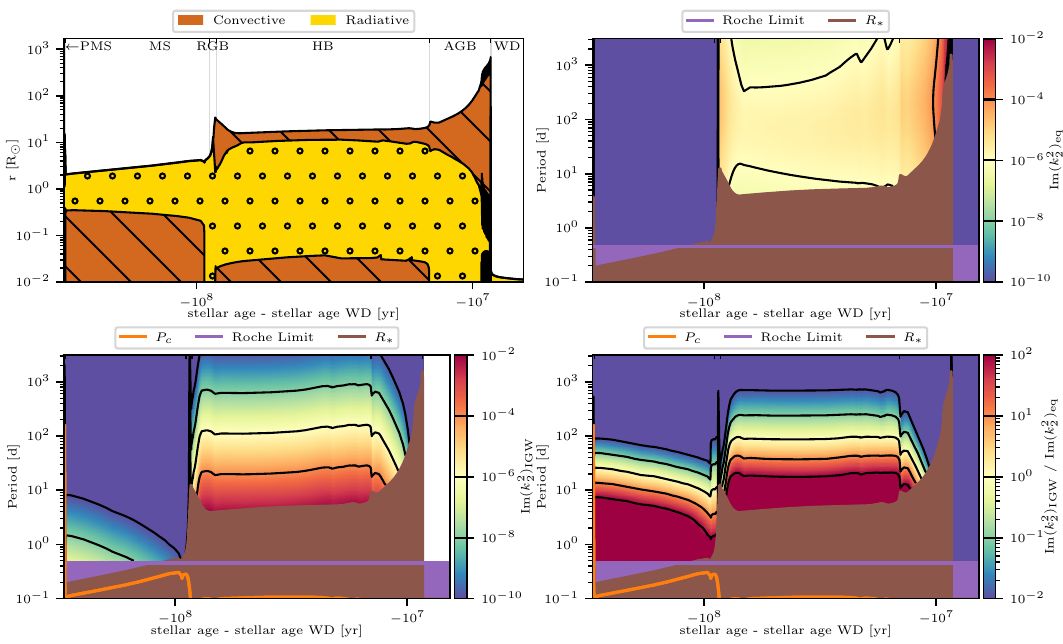}
            \caption{Same as Fig.~\ref{fig:FullM20}, but for a $M_\text{ZAMS}=3.5$ M$_\odot$ star.}\label{fig:FullM35}
        \end{figure*}
        \begin{figure*}
            \centering
            \includegraphics[width=\linewidth]{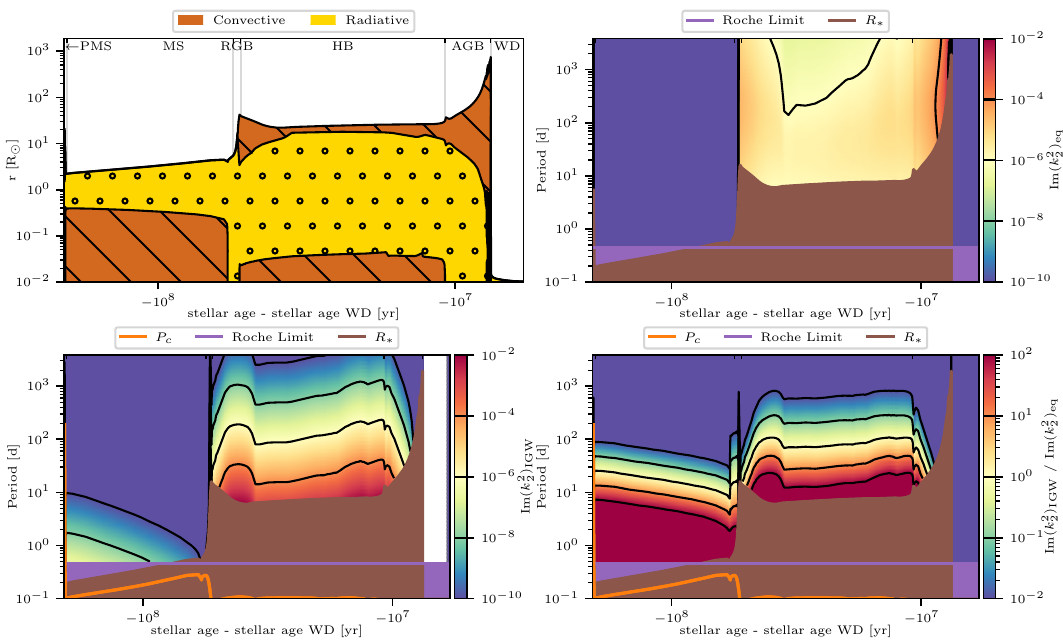}
            \caption{Same as Fig.~\ref{fig:FullM20}, but for a $M_\text{ZAMS}=4$ M$_\odot$ star.}\label{fig:FullM40}
        \end{figure*}
            The dependence of the tidal dissipation on the secondary mass is complex, and is dependent on whether we assume the orbital period and orbital distance to vary as a function of this parameter or not. Let's first assume both the orbital period and distance to be constant, a good approximation when only varying companion mass, which is negligible compared to the mass of the primary star as in the case of the planet. Then, the relation between the semi-major axis and the orbital period has only a weak dependence on the secondary mass. In this situation, there is a dependence on the secondary mass via $\varphi_T$ and $D_l$, with $\varphi_T$ being proportional to $M_2$ and $D_l$ being proportional to $M_2^2$. Therefore, the equilibrium tide dissipation given in Eq.~\eqref{eq:EquilibriumTide} is independent of the mass of the companion. For the dynamical tide dissipation (Eq.~\ref{eq:IGW}), there is direct dependence on the secondary mass, as well as in $\mathcal{F}$ which contains $\varphi_T$. The direct dependence on the mass of the companion cancels out again. This is expected as the Love number is defined as the ratio between the perturbation of the primary's gravitational potential induced by the presence of the companion and the tidal potential, evaluated at the stellar surface, and the direct dependence of the secondary mass in both the numerator and denominator is the same, resulting in a constant Love number as a function of $M_2$.
            
            However when the companion becomes more massive, the relation between the orbital period and the orbital distance ($a^3/P^2 =G(M_1+M_2)/4\pi^2$) becomes dependent on the secondary mass. Let's now consider the situation where the orbital period is fixed. Then, the change in the semi-major axis is proportional to $\sqrt[3]{M_1 + M_2}$ for this fixed orbital period. Given the fact that the tidal forcing $\mathcal{F}$ is inversely dependent on the semi-major axis cubed $\mathcal{F} \propto M_2/a^{3}$ (due to its dependance on $\varphi_T$; see Eq.~\ref{eq:TidalPotential}), the tidal forcing changes with a factor $\mathcal{F} \propto M_1M_2/(M_1 + M_2)$ compared to a companion with a small (negligible) mass. The dynamical tidal dissipation however also has a direct dependence on the semi-major axis (as well as on $M_2$). This dependence is exactly the same as the tidal forcing, and thus the dynamical tide dissipation will be constant with respect to changes in the secondary mass when the orbital period is fixed. The same is true for the equilibrium tide dissipation, as the dependence of the semi-major axis is the same in the tidal displacement and the tidal potential. This confirms that tidal dissipation is an intrisic property of a celestial body, which does not depend on the property of the secondary (excepting through the tidal frequency).
            
            On the other hand when fixing the semi-major axis $a$, but allowing the orbital period to vary by changing the secondary mass, the orbital frequency and thus the tidal frequency will change, with the orbital period being proportional to $1/\sqrt{M_1 + M_2}$. Since the semi-major axis remains constant in this situation, the tidal forcing becomes linear as a function of $M_2$ (cancelled out by the direct dependance of $\Im (k^2_2)_\text{IGW}$ on $M_2$). Therefore the dynamical tide dissipation will be proportional to $\omega_t^{8/3}$ and thus proportional to $P_\text{orb}^{-8/3}$. For higher mass companions, the dynamical tide dissipation will become stronger with a factor $((M_1 + M_2)/M_1)^{4/3}$ ($\Im (k^2_2)_\text{IGW} \propto ((M_1 + M_2)/M_1)^{4/3}$) compared to a companion with a small (negligible) mass. In this case the equilibrium tide dissipation will be affected as well. Due to the complex dependence of the turbulent viscosity on the tidal frequency, there is no direct proportionality constant that can be applied. However one can still read Fig.~\ref{fig:Imk2Eq} the same way, only for an orbital period that changes proportional to $1/\sqrt{M_1 + M_2}$. Depending on the orbital separation, and thus the regime of the turbulent viscosity, the equilibrium tide can become stronger or weaker (see Sect.~\ref{sec:EquilibriumTide}).\vspace*{-7pt}

        \subsection{Dependence of tidal dissipation on primary mass}\label{sec:TidalDissipationPrimaryMass}
            The strength of the dissipation of the equilibrium and dynamical tides are dependent on the mass of the primary star and the evolutionary stage. In this section we investigate the dependence of the tidal dissipation on the primary stellar mass and age using the grid of stellar models with initial masses between 1 and 4 M$_\odot$. The Kippenhahn diagram, together with the complex part of the Love number for both the equilibrium and dynamical tides as well as the ratio of their dissipation as a function of stellar age and orbital period, are shown in Figs.~\ref{fig:FullM20}, \ref{fig:FullM35} and \ref{fig:FullM40}, as well as in App.~\ref{sec:Appendix}.\footnote{The Love numbers for these evolutionary models can be found on zenodo: \url{https://doi.org/10.5281/zenodo.11519523}.}

            \subsubsection{Changes during the MS and RGB}
                When stars start their evolution with different initial masses, the internal structure on the MS changes. Lower mass stars have a convective envelope and a radiative interior (like the 1 M$_\odot$ star). Higher mass stars have a convective core and a radiative envelope, like for instance a 2 M$_\odot$ star (see Fig.~\ref{fig:FullM20}). In between there is a transition zone where there is a true three-layer structure, with a convective core, a radiative envelope and a convective envelope. This transition zone is dependent on the stellar evolution parameters used, where in our models the three-layer structure is apparent in the 1.2 and 1.4 M$_\odot$ models (see Figs.~\ref{fig:FullM12}~and~\ref{fig:FullM14}). For higher mass stars with a radiative envelope (the star still has a tiny convective envelope at the surface of the star), the equilibrium tide dissipation is drastically reduced, as well as the dynamical tide dissipation arising from the tiny convective envelope. At this point, the dynamical tide arising from the convective core propagating in the radiative envelope becomes dominant for short orbital periods (in agreement with \citealp{Zahn1975,Zahn1977,Goldreich1989}). This dominance over the equilibrium tide dissipation is less pronounced compared to the 1 M$_\odot$ star. This dynamical tide arising from the convective core decreases during the MS, as the size of the convective core is reduced during this phase, while at the same time the radius increases. When these stars evolve into RGB stars, the convective envelope reappears, and the equilibrium tide dissipation from this region and the dynamical tide excited at the boundary between the radiative core and the convective envelope become dominant again. This results in a sudden increase in the ratio of the dissipation of the dynamical to equilibrium tide, where afterwards it decreases similarly to the 1 M$_\odot$ star.

            \subsubsection{Changes during the HB and AGB}
                When stars evolve from the RGB into the HB, they will go through a helium flash for low mass stars (up to $M_\text{ZAMS} = $ 2 M$_\odot$), while for intermediate mass stars (above 2 M$_\odot$) the burning of helium starts gradually. For low mass stars, this results in a sudden contraction of the star, and a sudden increase in the ratio of the dissipation of the dynamical to equilibrium tide (see Sect.~\ref{sec:TidalDissipation}). For intermediate mass stars, the star will slowly contract, resulting in a slow increase of this ratio.
                
                The length of the RGB phase is also dependent on the star's initial mass. For lower mass stars the RGB phase is more pronounced compared to higher mass stars. For lower mass stars (with masses lower than 2 M$_\odot$ where the star goes through a helium flash), the maximal radius reached during the RGB is roughly the same with increasing stellar mass. Due to the increase in stellar mass, the period a companion has when it would orbit at the radius of the star at the tip of the RGB will be lower. For higher mass stars (with masses higher than 2 M$_\odot$ with gradual helium ignition) the maximal radius reached during the RGB increases with increasing stellar mass, but the radius remains small compared to the stars with a helium flash. At the same time, because of the increase in mass, the dynamical tide dissipation also increases (see Eq.~\ref{eq:IGW}), making the dynamical tide more important during the HB phase for these stars. Because of these two effects, the dynamical tide dissipation is important for companions at larger orbital distances than the primary's radius reached at the tip of the RGB (see e.g. Fig.~\ref{fig:FullM20} and \ref{fig:FullM35}).

                When increasing further in mass to the most massive model of 4 M$_\odot$ (see Fig.~\ref{fig:FullM40}), the convective envelope goes through a similar evolution when the star is contracting gradually after the start of helium burning. Hence the dissipation from both the equilibrium and dynamical tides increases, with similar relative strengths as for the 3.5 M$_\odot$ model (Fig.~\ref{fig:FullM35}). When the star is fully contracted throughout the HB, the convective envelope is smaller for the 4 M$_\odot$ model, reducing both the equilibrium and dynamical tide dissipation during this phase, where the dynamical tide dissipation is affected the most, reducing its relative importance. This effect is similar to the shift of convective to radiative envelope during the MS for 1 to 1.4 M$_\odot$ models.

                During the AGB phase, the trends of the equilibrium and dynamical tidal dissipation remain similar for all masses. At the start of the AGB phase, the ratio of the dissipation of the dynamical to equilibrium tide starts to decrease due to the increase in the primary's radius, similar to the RGB phase. As the relative importance of the dynamical tide dissipation increases with increasing mass during the HB phase, the relative importance also increases at the start of the AGB phase, but its importance becomes negligible early in the AGB phase. In the late stages of the AGB phase, the star will undergo thermal pulses (periodic instability marked by the sudden ignition of helium in a thin shell around the carbon-oxygen core, leading to temporary structural and luminosity changes). This will alter the internal structure of the star, and therefore the tidal dissipation. During a thermal pulse the star will undergo helium shell flashes, temporarily increasing its stellar radius, to afterwards go back to its original state. Therefore the equilibrium tide dissipation will increase during such a pulse. The dynamical tide dissipation will change as well, but throughout all the models the dynamical tide dissipation is negligible compared to the equilibrium tide dissipation during these thermal pulses. 
                
\section{Conclusion}\label{sec:Conclusion}
    This study investigated the dissipation of the equilibrium and dynamical tides from progressive internal gravity waves (where the dynamical tide from inertial waves are estimated to be negligible) along the entire evolution of stars, with a strong focus on the evolved phases. A grid of stellar evolutionary models was created with initial mass between 1 and 4 M$_\odot$. This allows the investigation of different stellar evolutionary effects on the tidal dissipation, such as the difference between a convective or radiative envelope during the MS, or the difference between a helium flash or gradual helium burning. Using these models we investigated the dependence on the orbital period, the mass of the primary star, and the mass of the companion. For stars with a sufficiently low primary mass such that the star has a radiative core and a convective envelope in the MS, the dynamical tide dissipation is dominant for orbital periods shorter than 10 days, in agreement with \citealp{Terquem1998,Goodman1998,Barker2010,Ahuir2021a}, while the equilibrium tide dissipation dominates for orbital periods longer than 50 days. For stars with a sufficiently high initial mass that the star has a radiative envelope and a convective core in the MS, tidal dissipation is dominated by the dynamical tide originating from the core for short orbital periods, in agreement with \cite{Zahn1975}. For longer orbital periods, the equilibrium tide dissipation dominates. At the start of the RGB, the dynamical tide dissipation first increases due to the increase in the spatial extent of the convective envelope that leads to an increase in the tidal forcing. For stars that already have a convective envelope during the MS this is a gradual increase, while for stars that have a radiative envelope during the MS the increase is instantaneous. During the RGB, the importance of the dynamical tide dissipation decreases as the radius of the star increases. At the start of the HB, the dynamical tide dissipation becomes more important again, as the radius of the star decreases either instantaneously in a helium flash or steadily when helium burning starts gradually. During the HB the importance of the dynamical tide dissipation increases with increasing stellar mass. For sufficiently high mass (4 M$_\odot$) the importance of the dynamical tide dissipation decreases instead of increasing as the size of the convective envelope shrinks. During the AGB, the equilibrium tide dissipation becomes dominant again due to the increase in stellar radius, similar to the RGB phase. During the WD phase, as the star is fully radiative, our formalism for the dynamical tide does not apply. However, this might be different when the dynamical tidal dissipation from standing g-modes and f-modes \citep{Fuller2011,Fuller2012,Fuller2013,Fuller2014,Veras2019} are taken into account.
    
    In a next step, these outcomes can be included in calculating the rate of change of the orbital distance of planetary and stellar companions along stellar evolution. The numerical model of a coplanar circular star-planet system called ESPEM \citep{Benbakoura2019,Ahuir2021b} can be used to study the orbital evolution of a star-planet and star-star system. This model includes the tidal dissipation of the equilibrium tide dissipation as well as the dynamical tide arising from inertial waves propagating in convective envelopes. This code will be improved by including the dissipation of the dynamical tide arising from progressive internal gravity waves \citep[similar to][]{Lazovik2021,Lazovik2023}, necessary for systems containing an evolved star. Furthermore the rotation of the primary star is reduced due to magnetic breaking in ESPEM \citep{Benbakoura2019,Ahuir2021b}. In evolved stars the magnetic breaking becomes negligible compared to the torque arising from the strong mass loss during this phase, which is not included in ESPEM as of now. This can be included by introducing a mass losing outer shell \citep{Madappatt2016}, which depends on the current mass-loss rate of the star. The mass-loss rate of AGB stars can be dependent on the presence of a companion, but ongoing work is still determining how much \citep{Decin2020,Aydi2022}. This will allow the study of the orbital distance evolution of star-planet and star-star systems around evolved stars and the resulting companion orbital distance and occurrence rate as this has been done for solar-like stars during the MS in \cite{Garcia2023}.
    
    Furthermore, the orbital evolution of star-planet systems around evolved stars can be studied taking into account the last breakthrough obtained through asteroseismology. Asteroseismology is now revealing the presence of potentially strong magnetic fields in the core of red giants \citep{Li2022,Li2023,Deheuvels2023}. Such strong magnetic fields can deeply modify the propagation of mixed modes, which behave as gravity modes in the core of red giants \citep[e.g.][]{Fuller2015,Bugnet2021,Mathis2021,Li2022,Mathis2023,Rui2023,Rui2024} and can be potentially excited by tides. The effect of such potential modification of the propagation and dissipation of such tidal modes still requires a dedicated investigation, potentially altering the dynamics of planetary systems around these magnetic stars.

\begin{acknowledgements}
    The authors would like to thank Matthias Fabry, Hannah Brinkman, Timothy Van Reeth, and Pablo Marchant for their help and support in setting up the MESA stellar evolution models as well as Cl\'ement Baruteau for the useful discussions. The authors would also like to thank the anonymous referee for their constructive comments which helped to improve the quality of the paper. M. Esseldeurs, S. Mathis and L. Decin acknowledge support from the FWO grant G0B3823N. M. Esseldeurs and L. Decin acknowledge support from the FWO grant G099720N, the KU Leuven C1 excellence grant MAESTRO C16/17/007 and the KU Leuven IDN grant ESCHER IDN/19/028. S. Mathis acknowledges support from the PLATO CNES grant at CEA/DAp, from the Programme National de Planétologie (PNP-CNRS/INSU) and from the European Research Council through HORIZON ERC SyG Grant 4D-STAR 101071505. While partially funded by the European Union, views and opinions expressed are however those of the author only and do not necessarily reflect those of the European Union or the European Research Council. Neither the European Union nor the granting authority can be held responsible for them.
\end{acknowledgements}

\bibliographystyle{aa}
\bibliography{bibliography}

\begin{thebibliography}{99}
\expandafter\ifx\csname natexlab\endcsname\relax\def\natexlab#1{#1}\fi

\bibitem[{{Abramowitz} \& {Stegun}(1972)}]{Abramowitz1972}
{Abramowitz}, M. \& {Stegun}, I.~A. 1972, {Handbook of Mathematical Functions}

\bibitem[{{Aerts} {et~al.}(2010){Aerts}, {Christensen-Dalsgaard}, \& {Kurtz}}]{Aerts2010}
{Aerts}, C., {Christensen-Dalsgaard}, J., \& {Kurtz}, D.~W. 2010, {Asteroseismology}

\bibitem[{{Aerts} {et~al.}(2019){Aerts}, {Mathis}, \& {Rogers}}]{Aerts2019}
{Aerts}, C., {Mathis}, S., \& {Rogers}, T.~M. 2019, \araa, 57, 35

\bibitem[{{Ahuir} {et~al.}(2021{\natexlab{a}}){Ahuir}, {Mathis}, \& {Amard}}]{Ahuir2021a}
{Ahuir}, J., {Mathis}, S., \& {Amard}, L. 2021{\natexlab{a}}, \aap, 651, A3

\bibitem[{{Ahuir} {et~al.}(2021{\natexlab{b}}){Ahuir}, {Strugarek}, {Brun}, \& {Mathis}}]{Ahuir2021b}
{Ahuir}, J., {Strugarek}, A., {Brun}, A.~S., \& {Mathis}, S. 2021{\natexlab{b}}, \aap, 650, A126

\bibitem[{{Alvan} {et~al.}(2015){Alvan}, {Strugarek}, {Brun}, {Mathis}, \& {Garcia}}]{Alvan2015}
{Alvan}, L., {Strugarek}, A., {Brun}, A.~S., {Mathis}, S., \& {Garcia}, R.~A. 2015, \aap, 581, A112

\bibitem[{{Amard} {et~al.}(2016){Amard}, {Palacios}, {Charbonnel}, {Gallet}, \& {Bouvier}}]{Amard2016}
{Amard}, L., {Palacios}, A., {Charbonnel}, C., {Gallet}, F., \& {Bouvier}, J. 2016, \aap, 587, A105

\bibitem[{{Asplund} {et~al.}(2009){Asplund}, {Grevesse}, {Sauval}, \& {Scott}}]{Asplund2009}
{Asplund}, M., {Grevesse}, N., {Sauval}, A.~J., \& {Scott}, P. 2009, \araa, 47, 481

\bibitem[{{Auclair Desrotour} {et~al.}(2015){Auclair Desrotour}, {Mathis}, \& {Le Poncin-Lafitte}}]{AuclairDesrotour2015}
{Auclair Desrotour}, P., {Mathis}, S., \& {Le Poncin-Lafitte}, C. 2015, \aap, 581, A118

\bibitem[{{Aydi} \& {Mohamed}(2022)}]{Aydi2022}
{Aydi}, E. \& {Mohamed}, S. 2022, \mnras, 513, 4405

\bibitem[{{Barker}(2020)}]{Barker2020}
{Barker}, A.~J. 2020, \mnras, 498, 2270

\bibitem[{{Barker} \& {Ogilvie}(2010)}]{Barker2010}
{Barker}, A.~J. \& {Ogilvie}, G.~I. 2010, \mnras, 404, 1849

\bibitem[{{Beck} {et~al.}(2024){Beck}, {Grossmann}, {Steinwender}, {Schimak}, {Muntean}, {Vrard}, {Patton}, {Merc}, {Mathur}, {Garcia}, {Pinsonneault}, {Rowan}, {Gaulme}, {Allende Prieto}, {Arellano-C{\'o}rdova}, {Cao}, {Corsaro}, {Creevey}, {Hambleton}, {Hanslmeier}, {Holl}, {Johnson}, {Mathis}, {Godoy-Rivera}, {S{\'\i}mon-D{\'\i}az}, \& {Zinn}}]{Beck2024}
{Beck}, P.~G., {Grossmann}, D.~H., {Steinwender}, L., {et~al.} 2024, \aap, 682, A7

\bibitem[{{Beck} {et~al.}(2018){Beck}, {Mathis}, {Gallet}, {Charbonnel}, {Benbakoura}, {Garc{\'\i}a}, \& {do Nascimento}}]{Beck2018}
{Beck}, P.~G., {Mathis}, S., {Gallet}, F., {et~al.} 2018, \mnras, 479, L123

\bibitem[{{Beck} {et~al.}(2022){Beck}, {Mathur}, {Hambleton}, {Garc{\'\i}a}, {Steinwender}, {Eisner}, {do Nascimento}, {Gaulme}, \& {Mathis}}]{Beck2022}
{Beck}, P.~G., {Mathur}, S., {Hambleton}, K., {et~al.} 2022, \aap, 667, A31

\bibitem[{{Benbakoura} {et~al.}(2019){Benbakoura}, {R{\'e}ville}, {Brun}, {Le Poncin-Lafitte}, \& {Mathis}}]{Benbakoura2019}
{Benbakoura}, M., {R{\'e}ville}, V., {Brun}, A.~S., {Le Poncin-Lafitte}, C., \& {Mathis}, S. 2019, \aap, 621, A124

\bibitem[{{Bl{\"o}cker}(1995)}]{Blocker1995}
{Bl{\"o}cker}, T. 1995, \aap, 297, 727

\bibitem[{{Bolmont} {et~al.}(2017){Bolmont}, {Gallet}, {Mathis}, {Charbonnel}, {Amard}, \& {Alibert}}]{Bolmont2017}
{Bolmont}, E., {Gallet}, F., {Mathis}, S., {et~al.} 2017, \aap, 604, A113

\bibitem[{{Bolmont} \& {Mathis}(2016)}]{Bolmont2016}
{Bolmont}, E. \& {Mathis}, S. 2016, Celestial Mechanics and Dynamical Astronomy, 126, 275

\bibitem[{{Bugnet} {et~al.}(2021){Bugnet}, {Prat}, {Mathis}, {Astoul}, {Augustson}, {Garc{\'\i}a}, {Mathur}, {Amard}, \& {Neiner}}]{Bugnet2021}
{Bugnet}, L., {Prat}, V., {Mathis}, S., {et~al.} 2021, \aap, 650, A53

\bibitem[{{Ceillier} {et~al.}(2017){Ceillier}, {Tayar}, {Mathur}, {Salabert}, {Garc{\'\i}a}, {Stello}, {Pinsonneault}, {van Saders}, {Beck}, \& {Bloemen}}]{Ceillier2017}
{Ceillier}, T., {Tayar}, J., {Mathur}, S., {et~al.} 2017, \aap, 605, A111

\bibitem[{{Cinquegrana} \& {Joyce}(2022)}]{Cinquegrana2022}
{Cinquegrana}, G.~C. \& {Joyce}, M. 2022, Research Notes of the American Astronomical Society, 6, 77

\bibitem[{{Decin} {et~al.}(2020){Decin}, {Montarg{\`e}s}, {Richards}, {Gottlieb}, {Homan}, {McDonald}, {El Mellah}, {Danilovich}, {Wallstr{\"o}m}, {Zijlstra}, {Baudry}, {Bolte}, {Cannon}, {De Beck}, {De Ceuster}, {de Koter}, {De Ridder}, {Etoka}, {Gobrecht}, {Gray}, {Herpin}, {Jeste}, {Lagadec}, {Kervella}, {Khouri}, {Menten}, {Millar}, {M{\"u}ller}, {Plane}, {Sahai}, {Sana}, {Van de Sande}, {Waters}, {Wong}, \& {Yates}}]{Decin2020}
{Decin}, L., {Montarg{\`e}s}, M., {Richards}, A.~M.~S., {et~al.} 2020, Science, 369, 1497

\bibitem[{{Deheuvels} {et~al.}(2023){Deheuvels}, {Li}, {Ballot}, \& {Ligni{\`e}res}}]{Deheuvels2023}
{Deheuvels}, S., {Li}, G., {Ballot}, J., \& {Ligni{\`e}res}, F. 2023, \aap, 670, L16

\bibitem[{{Dhouib} {et~al.}(2024){Dhouib}, {Baruteau}, {Mathis}, {Debras}, {Astoul}, \& {Rieutord}}]{Dhouib2024}
{Dhouib}, H., {Baruteau}, C., {Mathis}, S., {et~al.} 2024, \aap, 682, A85

\bibitem[{{D{\"o}llinger} \& {Hartmann}(2021)}]{Dollinger2021}
{D{\"o}llinger}, M.~P. \& {Hartmann}, M. 2021, \apjs, 256, 10

\bibitem[{{Duguid} {et~al.}(2020){Duguid}, {Barker}, \& {Jones}}]{Duguid2020}
{Duguid}, C.~D., {Barker}, A.~J., \& {Jones}, C.~A. 2020, \mnras, 497, 3400

\bibitem[{{Fuller}(2017)}]{Fuller2017}
{Fuller}, J. 2017, \mnras, 472, 1538

\bibitem[{{Fuller} {et~al.}(2015){Fuller}, {Cantiello}, {Stello}, {Garcia}, \& {Bildsten}}]{Fuller2015}
{Fuller}, J., {Cantiello}, M., {Stello}, D., {Garcia}, R.~A., \& {Bildsten}, L. 2015, Science, 350, 423

\bibitem[{{Fuller} \& {Lai}(2011)}]{Fuller2011}
{Fuller}, J. \& {Lai}, D. 2011, \mnras, 412, 1331

\bibitem[{{Fuller} \& {Lai}(2012)}]{Fuller2012}
{Fuller}, J. \& {Lai}, D. 2012, \mnras, 421, 426

\bibitem[{{Fuller} \& {Lai}(2013)}]{Fuller2013}
{Fuller}, J. \& {Lai}, D. 2013, \mnras, 430, 274

\bibitem[{{Fuller} \& {Lai}(2014)}]{Fuller2014}
{Fuller}, J. \& {Lai}, D. 2014, \mnras, 444, 3488

\bibitem[{{Fuller} {et~al.}(2019){Fuller}, {Piro}, \& {Jermyn}}]{Fuller2019}
{Fuller}, J., {Piro}, A.~L., \& {Jermyn}, A.~S. 2019, \mnras, 485, 3661

\bibitem[{{Gallet} {et~al.}(2017){Gallet}, {Bolmont}, {Mathis}, {Charbonnel}, \& {Amard}}]{Gallet2017}
{Gallet}, F., {Bolmont}, E., {Mathis}, S., {Charbonnel}, C., \& {Amard}, L. 2017, \aap, 604, A112

\bibitem[{{Garc{\'\i}a} {et~al.}(2023){Garc{\'\i}a}, {Gourv{\`e}s}, {Santos}, {Strugarek}, {Godoy-Rivera}, {Mathur}, {Delsanti}, {Breton}, {Beck}, {Brun}, \& {Mathis}}]{Garcia2023}
{Garc{\'\i}a}, R.~A., {Gourv{\`e}s}, C., {Santos}, A.~R.~G., {et~al.} 2023, \aap, 679, L12

\bibitem[{{Garc{\'\i}a-Segura} {et~al.}(2016){Garc{\'\i}a-Segura}, {Villaver}, {Manchado}, {Langer}, \& {Yoon}}]{Garcia-Segura2016}
{Garc{\'\i}a-Segura}, G., {Villaver}, E., {Manchado}, A., {Langer}, N., \& {Yoon}, S.~C. 2016, \apj, 823, 142

\bibitem[{{Goldreich} \& {Nicholson}(1989)}]{Goldreich1989}
{Goldreich}, P. \& {Nicholson}, P.~D. 1989, \apj, 342, 1079

\bibitem[{{Goodman} \& {Dickson}(1998)}]{Goodman1998}
{Goodman}, J. \& {Dickson}, E.~S. 1998, \apj, 507, 938

\bibitem[{{Henyey} {et~al.}(1965){Henyey}, {Vardya}, \& {Bodenheimer}}]{Henyey1965}
{Henyey}, L., {Vardya}, M.~S., \& {Bodenheimer}, P. 1965, \apj, 142, 841

\bibitem[{{Hon} {et~al.}(2023){Hon}, {Huber}, {Rui}, {Fuller}, {Veras}, {Kuszlewicz}, {Kochukhov}, {Stokholm}, {R{\o}rsted}, {Y{\i}ld{\i}z}, {Orhan}, {{\"O}rtel}, {Jiang}, {Hey}, {Isaacson}, {Zhang}, {Vrard}, {Stassun}, {Shappee}, {Tayar}, {Claytor}, {Beard}, {Bedding}, {Brinkman}, {Campante}, {Chaplin}, {Chontos}, {Giacalone}, {Holcomb}, {Howard}, {Lubin}, {MacDougall}, {Montet}, {Murphy}, {Ong}, {Pidhorodetska}, {Polanski}, {Rice}, {Stello}, {Tyler}, {Van Zandt}, \& {Weiss}}]{Hon2023}
{Hon}, M., {Huber}, D., {Rui}, N.~Z., {et~al.} 2023, \nat, 618, 917

\bibitem[{{Jermyn} {et~al.}(2023){Jermyn}, {Bauer}, {Schwab}, {Farmer}, {Ball}, {Bellinger}, {Dotter}, {Joyce}, {Marchant}, {Mombarg}, {Wolf}, {Sunny Wong}, {Cinquegrana}, {Farrell}, {Smolec}, {Thoul}, {Cantiello}, {Herwig}, {Toloza}, {Bildsten}, {Townsend}, \& {Timmes}}]{Jermyn2023}
{Jermyn}, A.~S., {Bauer}, E.~B., {Schwab}, J., {et~al.} 2023, \apjs, 265, 15

\bibitem[{{Kaula}(1962)}]{Kaula1962}
{Kaula}, W.~M. 1962, \aj, 67, 300

\bibitem[{{Kawaler}(1988)}]{Kawaler1988}
{Kawaler}, S.~D. 1988, \apj, 333, 236

\bibitem[{{Lau} {et~al.}(2022){Lau}, {Cantiello}, {Jermyn}, {MacLeod}, {Mandel}, \& {Price}}]{Lau2022}
{Lau}, M. Y.~M., {Cantiello}, M., {Jermyn}, A.~S., {et~al.} 2022, arXiv e-prints, arXiv:2210.15848

\bibitem[{{Lazovik}(2021)}]{Lazovik2021}
{Lazovik}, Y.~A. 2021, \mnras, 508, 3408

\bibitem[{{Lazovik}(2023)}]{Lazovik2023}
{Lazovik}, Y.~A. 2023, \mnras, 520, 3749

\bibitem[{{Lee} {et~al.}(2023){Lee}, {Do}, {Park}, {Lim}, {Choi}, {Koo}, {Bang}, {Oh}, {Han}, \& {Chang}}]{Lee2023}
{Lee}, B.-C., {Do}, H.-J., {Park}, M.-G., {et~al.} 2023, \aap, 678, A106

\bibitem[{{Li} {et~al.}(2022){Li}, {Deheuvels}, {Ballot}, \& {Ligni{\`e}res}}]{Li2022}
{Li}, G., {Deheuvels}, S., {Ballot}, J., \& {Ligni{\`e}res}, F. 2022, \nat, 610, 43

\bibitem[{{Li} {et~al.}(2023){Li}, {Deheuvels}, {Li}, {Ballot}, \& {Ligni{\`e}res}}]{Li2023}
{Li}, G., {Deheuvels}, S., {Li}, T., {Ballot}, J., \& {Ligni{\`e}res}, F. 2023, \aap, 680, A26

\bibitem[{{Love}(1911)}]{Love1911}
{Love}, A.~E.~H. 1911, {Some Problems of Geodynamics}

\bibitem[{{Madappatt} {et~al.}(2016){Madappatt}, {De Marco}, \& {Villaver}}]{Madappatt2016}
{Madappatt}, N., {De Marco}, O., \& {Villaver}, E. 2016, \mnras, 463, 1040

\bibitem[{{Marigo} \& {Aringer}(2009)}]{Marigo2009}
{Marigo}, P. \& {Aringer}, B. 2009, \aap, 508, 1539

\bibitem[{{Mathis}(2013)}]{Mathis2013}
{Mathis}, S. 2013, in Lecture Notes in Physics, Berlin Springer Verlag, ed. M.~{Goupil}, K.~{Belkacem}, C.~{Neiner}, F.~{Ligni{\`e}res}, \& J.~J. {Green}, Vol. 865, 23

\bibitem[{{Mathis}(2015)}]{Mathis2015}
{Mathis}, S. 2015, \aap, 580, L3

\bibitem[{{Mathis}(2018)}]{Mathis2018}
{Mathis}, S. 2018, in Handbook of Exoplanets, ed. H.~J. {Deeg} \& J.~A. {Belmonte}, 24

\bibitem[{{Mathis} {et~al.}(2016){Mathis}, {Auclair-Desrotour}, {Guenel}, {Gallet}, \& {Le Poncin-Lafitte}}]{Mathis2016}
{Mathis}, S., {Auclair-Desrotour}, P., {Guenel}, M., {Gallet}, F., \& {Le Poncin-Lafitte}, C. 2016, \aap, 592, A33

\bibitem[{{Mathis} \& {Bugnet}(2023)}]{Mathis2023}
{Mathis}, S. \& {Bugnet}, L. 2023, \aap, 676, L9

\bibitem[{{Mathis} {et~al.}(2021){Mathis}, {Bugnet}, {Prat}, {Augustson}, {Mathur}, \& {Garcia}}]{Mathis2021}
{Mathis}, S., {Bugnet}, L., {Prat}, V., {et~al.} 2021, \aap, 647, A122

\bibitem[{{Mathis} \& {Le Poncin-Lafitte}(2009)}]{Mathis2009}
{Mathis}, S. \& {Le Poncin-Lafitte}, C. 2009, \aap, 497, 889

\bibitem[{{McDonald} \& {Zijlstra}(2015)}]{McDonald2015}
{McDonald}, I. \& {Zijlstra}, A.~A. 2015, \mnras, 448, 502

\bibitem[{{Mustill} \& {Villaver}(2012)}]{Mustill2012}
{Mustill}, A.~J. \& {Villaver}, E. 2012, \apj, 761, 121

\bibitem[{{Nowak} {et~al.}(2013){Nowak}, {Niedzielski}, {Wolszczan}, {Adam{\'o}w}, \& {Maciejewski}}]{Nowak2013}
{Nowak}, G., {Niedzielski}, A., {Wolszczan}, A., {Adam{\'o}w}, M., \& {Maciejewski}, G. 2013, \apj, 770, 53

\bibitem[{{Ogilvie}(2013)}]{Ogilvie2013}
{Ogilvie}, G.~I. 2013, \mnras, 429, 613

\bibitem[{{Ogilvie}(2014)}]{Ogilvie2014}
{Ogilvie}, G.~I. 2014, \araa, 52, 171

\bibitem[{{Ogilvie} \& {Lin}(2004)}]{Ogilvie2004}
{Ogilvie}, G.~I. \& {Lin}, D.~N.~C. 2004, \apj, 610, 477

\bibitem[{{Ogilvie} \& {Lin}(2007)}]{Ogilvie2007}
{Ogilvie}, G.~I. \& {Lin}, D.~N.~C. 2007, \apj, 661, 1180

\bibitem[{{Oomen} {et~al.}(2018){Oomen}, {Van Winckel}, {Pols}, {Nelemans}, {Escorza}, {Manick}, {Kamath}, \& {Waelkens}}]{Oomen2018}
{Oomen}, G.-M., {Van Winckel}, H., {Pols}, O., {et~al.} 2018, \aap, 620, A85

\bibitem[{{Paxton} {et~al.}(2011){Paxton}, {Bildsten}, {Dotter}, {Herwig}, {Lesaffre}, \& {Timmes}}]{Paxton2011}
{Paxton}, B., {Bildsten}, L., {Dotter}, A., {et~al.} 2011, \apjs, 192, 3

\bibitem[{{Paxton} {et~al.}(2013){Paxton}, {Cantiello}, {Arras}, {Bildsten}, {Brown}, {Dotter}, {Mankovich}, {Montgomery}, {Stello}, {Timmes}, \& {Townsend}}]{Paxton2013}
{Paxton}, B., {Cantiello}, M., {Arras}, P., {et~al.} 2013, \apjs, 208, 4

\bibitem[{{Paxton} {et~al.}(2015){Paxton}, {Marchant}, {Schwab}, {Bauer}, {Bildsten}, {Cantiello}, {Dessart}, {Farmer}, {Hu}, {Langer}, {Townsend}, {Townsley}, \& {Timmes}}]{Paxton2015}
{Paxton}, B., {Marchant}, P., {Schwab}, J., {et~al.} 2015, \apjs, 220, 15

\bibitem[{{Paxton} {et~al.}(2018){Paxton}, {Schwab}, {Bauer}, {Bildsten}, {Blinnikov}, {Duffell}, {Farmer}, {Goldberg}, {Marchant}, {Sorokina}, {Thoul}, {Townsend}, \& {Timmes}}]{Paxton2018}
{Paxton}, B., {Schwab}, J., {Bauer}, E.~B., {et~al.} 2018, \apjs, 234, 34

\bibitem[{{Paxton} {et~al.}(2019){Paxton}, {Smolec}, {Schwab}, {Gautschy}, {Bildsten}, {Cantiello}, {Dotter}, {Farmer}, {Goldberg}, {Jermyn}, {Kanbur}, {Marchant}, {Thoul}, {Townsend}, {Wolf}, {Zhang}, \& {Timmes}}]{Paxton2019}
{Paxton}, B., {Smolec}, R., {Schwab}, J., {et~al.} 2019, \apjs, 243, 10

\bibitem[{{Pereira} {et~al.}(2024){Pereira}, {Grunblatt}, {Psaridi}, {Campante}, {Cunha}, {Santos}, {Bossini}, {Thorngren}, {Hellier}, {Bouchy}, {Lendl}, {Mounzer}, {Udry}, {Beard}, {Brinkman}, {Isaacson}, {Quinn}, {Tyler}, {Zhou}, {Howell}, {Howard}, {Jenkins}, {Seager}, {Vanderspek}, {Winn}, {Saunders}, \& {Huber}}]{Pereira2024}
{Pereira}, F., {Grunblatt}, S.~K., {Psaridi}, A., {et~al.} 2024, \mnras, 527, 6332

\bibitem[{{Reimers}(1975)}]{Reimers1975}
{Reimers}, D. 1975, in Problems in stellar atmospheres and envelopes., 229--256

\bibitem[{{Remus} {et~al.}(2012){Remus}, {Mathis}, \& {Zahn}}]{Remus2012}
{Remus}, F., {Mathis}, S., \& {Zahn}, J.~P. 2012, \aap, 544, A132

\bibitem[{{Rieutord}(2015)}]{Rieutord2015}
{Rieutord}, M. 2015, {Fluid Dynamics: An Introduction}

\bibitem[{{Rui} \& {Fuller}(2023)}]{Rui2023}
{Rui}, N.~Z. \& {Fuller}, J. 2023, \mnras, 523, 582

\bibitem[{{Rui} {et~al.}(2024){Rui}, {Ong}, \& {Mathis}}]{Rui2024}
{Rui}, N.~Z., {Ong}, J.~M.~J., \& {Mathis}, S. 2024, \mnras, 527, 6346

\bibitem[{{Sato} {et~al.}(2008){Sato}, {Izumiura}, {Toyota}, {Kambe}, {Ikoma}, {Omiya}, {Masuda}, {Takeda}, {Murata}, {Itoh}, {Ando}, {Yoshida}, {Kokubo}, \& {Ida}}]{Sato2008}
{Sato}, B., {Izumiura}, H., {Toyota}, E., {et~al.} 2008, \pasj, 60, 539

\bibitem[{{Saunders} {et~al.}(2022){Saunders}, {Grunblatt}, {Huber}, {Collins}, {Jensen}, {Vanderburg}, {Brahm}, {Jord{\'a}n}, {Espinoza}, {Henning}, {Hobson}, {Quinn}, {Zhou}, {Butler}, {Crause}, {Kuhn}, {Moses Mogotsi}, {Hellier}, {Angus}, {Hattori}, {Chontos}, {Ricker}, {Jenkins}, {Tenenbaum}, {Latham}, {Seager}, {Vanderspek}, {Winn}, {Stockdale}, \& {Cloutier}}]{Saunders2022}
{Saunders}, N., {Grunblatt}, S.~K., {Huber}, D., {et~al.} 2022, \aj, 163, 53

\bibitem[{{Siess} {et~al.}(2000){Siess}, {Dufour}, \& {Forestini}}]{Siess2000}
{Siess}, L., {Dufour}, E., \& {Forestini}, M. 2000, \aap, 358, 593

\bibitem[{{Skumanich}(1972)}]{Skumanich1972}
{Skumanich}, A. 1972, \apj, 171, 565

\bibitem[{{Terquem} {et~al.}(1998){Terquem}, {Papaloizou}, {Nelson}, \& {Lin}}]{Terquem1998}
{Terquem}, C., {Papaloizou}, J.~C.~B., {Nelson}, R.~P., \& {Lin}, D.~N.~C. 1998, \apj, 502, 788

\bibitem[{{Van Winckel}(2003)}]{VanWinckel2003}
{Van Winckel}, H. 2003, \araa, 41, 391

\bibitem[{{Veras} \& {Fuller}(2019)}]{Veras2019}
{Veras}, D. \& {Fuller}, J. 2019, \mnras, 489, 2941

\bibitem[{{Verbunt} \& {Phinney}(1995)}]{Verbunt1995}
{Verbunt}, F. \& {Phinney}, E.~S. 1995, \aap, 296, 709

\bibitem[{{Viallet} {et~al.}(2015){Viallet}, {Meakin}, {Prat}, \& {Arnett}}]{Viallet2015}
{Viallet}, M., {Meakin}, C., {Prat}, V., \& {Arnett}, D. 2015, \aap, 580, A61

\bibitem[{{Vlemmings} {et~al.}(2018){Vlemmings}, {Khouri}, {De Beck}, {Olofsson}, {Garc{\'\i}a-Segura}, {Villaver}, {Baudry}, {Humphreys}, {Maercker}, \& {Ramstedt}}]{Vlemmings2018}
{Vlemmings}, W.~H.~T., {Khouri}, T., {De Beck}, E., {et~al.} 2018, \aap, 613, L4

\bibitem[{{Weinberg} {et~al.}(2017){Weinberg}, {Sun}, {Arras}, \& {Essick}}]{Weinberg2017}
{Weinberg}, N.~N., {Sun}, M., {Arras}, P., \& {Essick}, R. 2017, \apjl, 849, L11

\bibitem[{{Witte} \& {Savonije}(2002)}]{Witte2002}
{Witte}, M.~G. \& {Savonije}, G.~J. 2002, \aap, 386, 222

\bibitem[{{Wu}(2005{\natexlab{a}})}]{Wu2005a}
{Wu}, Y. 2005{\natexlab{a}}, \apj, 635, 674

\bibitem[{{Wu}(2005{\natexlab{b}})}]{Wu2005b}
{Wu}, Y. 2005{\natexlab{b}}, \apj, 635, 688

\bibitem[{{Zahn}(1966)}]{Zahn1966a}
{Zahn}, J.~P. 1966, Annales d'Astrophysique, 29, 313

\bibitem[{{Zahn}(1975)}]{Zahn1975}
{Zahn}, J.~P. 1975, \aap, 41, 329

\bibitem[{{Zahn}(1977)}]{Zahn1977}
{Zahn}, J.~P. 1977, \aap, 57, 383

\bibitem[{{Zahn}(1989)}]{Zahn1989a}
{Zahn}, J.~P. 1989, \aap, 220, 112

\bibitem[{{Zahn}(1992)}]{Zahn1992}
{Zahn}, J.~P. 1992, \aap, 265, 115

\bibitem[{{Zahn} \& {Bouchet}(1989)}]{Zahn1989b}
{Zahn}, J.~P. \& {Bouchet}, L. 1989, \aap, 223, 112

\end{thebibliography}

\begin{appendix}
    \section{Ekman number in evolved stars}\label{sec:Ekman}
        \begin{figure}
            \centering
            \includegraphics[width=\linewidth]{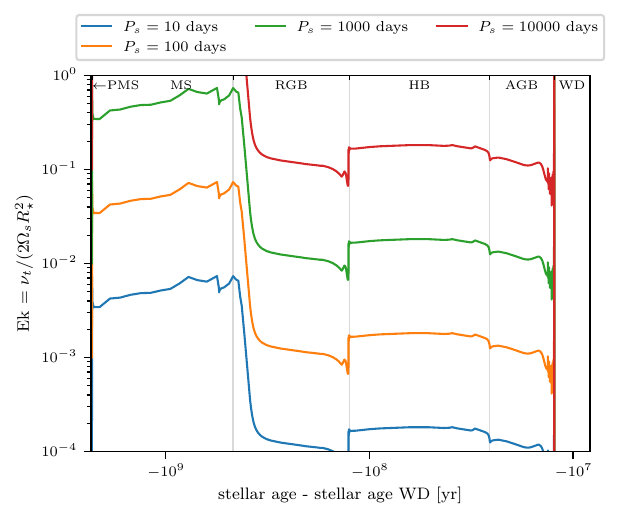}
            \vspace*{-10pt}
            \caption{The Ekman number calculated in the middle of the external convective envelope for different rotational periods during the lifetime of a 1.6 M$_\odot$ star. In brown the radius of the star is shown for reference.
            }\label{fig:Ekman}
        \end{figure}
        The Ekman number $\mathrm{Ek}=\nu_t/(2\Omega_s R_\star^2)$ is defined as the ratio of the viscous force to the Coriolis force (where the eddy viscosity is used, assuming that the action of turbulent convection on tidal inertial waves can be modelled as an effective viscous friction). It is a measure of the importance of the viscous dissipation in the star compared to its rotation, and is important for the damping of inertial waves and modes. If the Ekman number is high, the resonant dissipation of tidally-forced inertial waves is damped \citep[e.g.][]{Ogilvie2004,AuclairDesrotour2015} and is not an efficient channel for tidal dissipation. To correctly model the Ekman number, a correct estimation of the rotation rate of the star is necessary. Since we do not take into account the rotation of the star in our stellar models, or its potential spin-up from a massive companion, a true estimate of the Ekman number cannot be given. However, the Ekman number can be calculated assuming different (constant) rotation rates. \cite{Fuller2019} have computed the rotational period for a $M_\text{ZAMS}=1.6$ M$_\odot$ using MESA stellar models where they implemented the transport of angular momentum by magnetic stresses obtaining predictions that match asteroseismic measurements of stellar core rotation. They have shown that rotation varies during the evolved phases between 10 and 10$^4$ days, in agreement with observations of the surface rotation of this type of stars \citep{Ceillier2017,Vlemmings2018}. The Ekman number, evaluated in the middle of the convective envelope (which provides a reasonable order of magnitude estimate even if it varies with stellar radius; \citealp{Mathis2016}) for different rotational periods and a $M_\text{ZAMS}=1.6$ M$_\odot$ model, is shown in Fig.~\ref{fig:Ekman}. \cite{Fuller2019} predicts rotational periods during the RGB to start of the order of 10 days and quickly rise to about 10$^4$ days. Looking at Fig.~\ref{fig:Ekman}, at the beginning of the RGB, the Ekman number (for a rotational period of 10 days) is of the order of $10^{-2}$,  while it becomes of the order of $10^{-1}$ at the end of the RGB (for rotational period of 10$^4$ days). During the HB the rotational period in \cite{Fuller2019} decreases to be of the order of a few hundred days, which would result in an Ekman number of a few times $10^{-3}$. During the AGB, the rotational period increases again to be of the order of 10$^4$ days and higher, resulting in an Ekman number of a $10^{-1}$ and higher. This results in Ekman numbers is of the order of $10^{-2}$ to $10^{0}$ during the RGB and AGB, and a little bit lower during the HB. These values are sufficiently high to neglect the resonant dissipation of tidally-excited inertial modes.

    \section{Dynamical tidal dissipation}\label{sec:imk2dyn}
        The energy luminosity of the dissipation of tidally-excited progressive internal gravity waves has been derived by \cite{Ahuir2021a}. We refer the reader to this article for the details of the theoretical calculations leading to the following expression in the case of a star with a radiative core and a convective envelope:
        \begin{equation}\begin{aligned}
            L_\text{E, IGW} = -\frac{3^\frac{2}{3}\Gamma^2\left(\frac{1}{3}\right)}{8\pi} \omega_t^\frac{11}{3}&(l(l+1))^{-\frac{4}{3}}\\ &\rho_0(r_\text{int})r_\text{int}\left|\frac{\dd N^2}{\dd \ln r}\right|_{r_\text{int}}^{-\frac{1}{3}} \mathcal{F}^2\ ,
        \end{aligned}\end{equation}
        where $r_\text{int}$ is the radius of the radiative-convective interface. In a circular coplanar orbit, the energy luminosity can be expressed in terms of the complex part of the Love number as \citep{Ogilvie2014}:
        \begin{align}
            L_\text{E, IGW} &= -\omega_t \frac{(2l + 1) R_\star |\varphi_T(R_\star)|^2}{8\pi G}\Im\left(k_2^2\right)_\text{IGW}\\
            &= -\frac{3}{4} \omega_t GM_2^2\frac{R_\star^5}{a^6}\Im\left(k_2^2\right)_\text{IGW}\ .
        \end{align}
        These equations can be used to derive the expression for the complex part of the Love number for the dissipation of tidally-excited progressive internal gravity waves:
        \begin{equation}
            \begin{aligned}
                \Im\left(k_2^2\right)_\text{IGW} = \frac{3^{-\frac{1}{3}}\Gamma^2\left(\frac{1}{3}\right)}{2\pi} \omega_t^\frac{8}{3}(l(l+&1))^{-\frac{4}{3}} \frac{a^6}{GM_2^2R_\star^5}\\
                &\rho_0(r_\text{int})r_\text{int}\left|\frac{\dd N^2}{\dd \ln r}\right|_{r_\text{int}}^{-\frac{1}{3}} \mathcal{F}^2\ .
            \end{aligned}
        \end{equation}
        This expression can now be generalised to a three-layer structure (analogous to Sect.~4.3 of \citealp{Ahuir2021a}) to
        \begin{equation}
            \begin{aligned}
                \Im\left(k_2^2\right)_\text{IGW} = &\frac{3^{-\frac{1}{3}}\Gamma^2\left(\frac{1}{3}\right)}{2\pi} \omega_t^\frac{8}{3}(l(l+1))^{-\frac{4}{3}} \frac{a^6}{GM_2^2R_\star^5}\\
               &\begin{aligned}\times\Bigg(
                \rho_0\left(r_\text{in}\right) r_\text{in}&\left|\frac{\dd N^2}{\mathrm{~d} \ln r}\right|_{r_\text{in}}^{-\frac{1}{3}} \mathcal{F}_\text{{in}}^2\\
                &+\rho_0\left(r_\text{out}\right) r_\text{out}\left|\frac{\dd N^2}{\mathrm{~d} \ln r}\right|_{r_\text{out}}^{-\frac{1}{3}} \mathcal{F}_\text{{out}}^2
            \Bigg)\ .\end{aligned}
           \end{aligned}
        \end{equation}
\newpage

\section{Comparison to \cite{Ahuir2021a}}\label{sec:Comparison}
    To benchmark our implementation of the tidal dissipation prescriptions, as well as to estimate the robustness of the solutions for different stellar structure and evolution numerical codes, we compare our results to the results obtained by \cite{Ahuir2021a} who used stellar models with initial masses between 0.4 and 1.4 M$_\odot$ using the stellar evolution code STAREVOL \citep{Siess2000,Amard2016}.
    
\begin{figure}
        \centering
        \includegraphics[width=\linewidth]{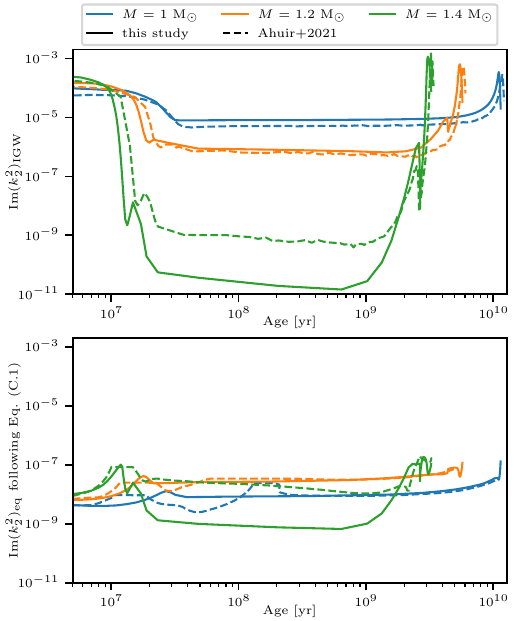}
        \caption{Complex part of the Love number for dynamical ($\Im (k^2_2)_\text{IGW}$; upper plot) and equilibrium ($\Im (k^2_2)_\text{eq}$ following Eq.~\ref{eq:EquilibriumTideAhuir2021}; lower plot) tides, as a function of stellar age, for both \cite{Ahuir2021a} in the dashed lines and this study in the solid lines.\\}\label{fig:Ahuir2021}

        \centering
        \includegraphics[width=\linewidth]{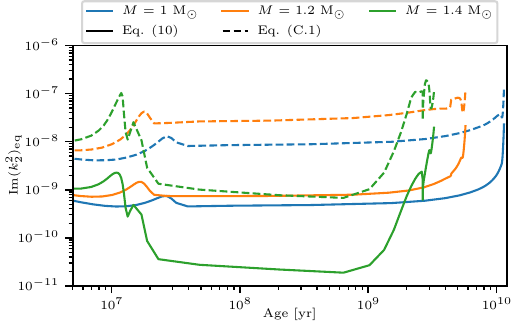}
        \caption{Complex part of the Love number for equilibrium tide ($\Im (k^2_2)_\text{eq}$), as a function of stellar age, for both the \cite{Ahuir2021a} formalism (Eq.~\ref{eq:EquilibriumTideAhuir2021}) and this study (Eq.~\ref{eq:EquilibriumTide}).}\label{fig:Ahuir2021Equilibrium}
    \end{figure}

    In \cite{Ahuir2021a}, the dissipation of the equilibrium tide is calculated assuming a thin convective envelope, assuming a constant density in the convective envelope, and approximating the mixing length by its maximum, such that the equilibrium tide dissipation becomes:
    \begin{equation}\label{eq:EquilibriumTideAhuir2021}
        \Im (k^2_2)_\mathrm{eq} = \frac{696}{35} \left|t_c \omega_t\right|\frac{R_\star}{G M_1} V_{\mathrm{c}}^2 (1-\beta)\frac{1-\alpha^9}{1-\alpha^3}F(\omega_t)\ ,
    \end{equation}
    where $V_{\mathrm{c}}$ is the convective velocity, $\alpha$ and $\beta$ are the ratio of the convective to the total radius and mass respectively, and $F(\omega_t)$ is the dependence of the turbulent viscosity on the tidal frequency (see Eq.~\ref{eq:TurbulentViscosity}). Assuming this formalism, the equilibrium tide is compared to the one calculated in \cite{Ahuir2021a} in Fig.~\ref{fig:Ahuir2021} (top panel). Here the dissipation is consistent between the MESA and STAREVOL models, except for the 1.4 M$_\odot$ model where the dissipation is weaker in this study. This is due to the different opacity tables used in the two stellar evolution codes, resulting in lower densities in the small convective envelope.
    
    For evolved stars with a large convective envelope, these approximations are not valid anymore. Therefore, we compute here the equilibrium tide dissipation with taking into account realistic stellar profiles (see Sect.~\ref{sec:Equilibrium} and \citealp{Barker2020}). The two different formalisms are shown in Fig.~\ref{fig:Ahuir2021Equilibrium}. Here it can be seen that the equilibrium tide dissipation calculated in this study is significantly weaker than the one calculated in \cite{Ahuir2021a}. As more approximations were made to estimate the equilibrium tide dissipation in \cite{Ahuir2021a}, the dissipation of the equilibrium tide calculated in this study is more reliable.

    In Fig.~\ref{fig:Ahuir2021} the tidal dissipation from IGW is shown on the top for 1, 1.2 and 1.4 M$_\odot$ stars and a 1 M$_\text{Jup}$ companion on an orbit of 1 d, using both the STAREVOL models used in \cite{Ahuir2021a} as well as the MESA models used in this study. When performing this benchmark, a unit mistake was found in the numerical computations of \cite{Ahuir2021a}, resulting in a reduction of the dynamical tide dissipation by roughly an order of magnitude compared to what was reported by \cite{Ahuir2021a}. This has no significant impact on their conclusions. For the 1 and 1.2 M$_\odot$ models the results are almost identical when using MESA and STAREVOL models. For the 1.4 M$_\odot$ model, the dynamical tide dissipation are weaker in this study. This is because the star has a negligible convective envelope during the MS, and as the stellar evolution codes use different opacity tables, these regions are slightly different when using MESA or STAREVOL, resulting in different densities in these layers. During the further evolution of the stars, when the convective envelope grows again, the difference vanishes. Therefore the ratio of the dynamical to equilibrium tidal dissipation remains roughly the same when using MESA or STAREVOL models. The results are in good agreement with those of \cite{Terquem1998,Goodman1998,Barker2010}. This validates our implementation in MESA and show that the results are robust for different stellar structure and evolution codes.

\section{Tidal dissipation evaluation for complementary masses of the stellar evolution grid.}\label{sec:Appendix}
    This section contains the tidal dissipation figures for the remaining stellar evolutionary models not illustrated in the main text. Their main patterns have been identified and explained in Sect.~\ref{sec:TidalDissipationPrimaryMass}.

    \begin{figure*}
        \centering
        \includegraphics[width=\linewidth]{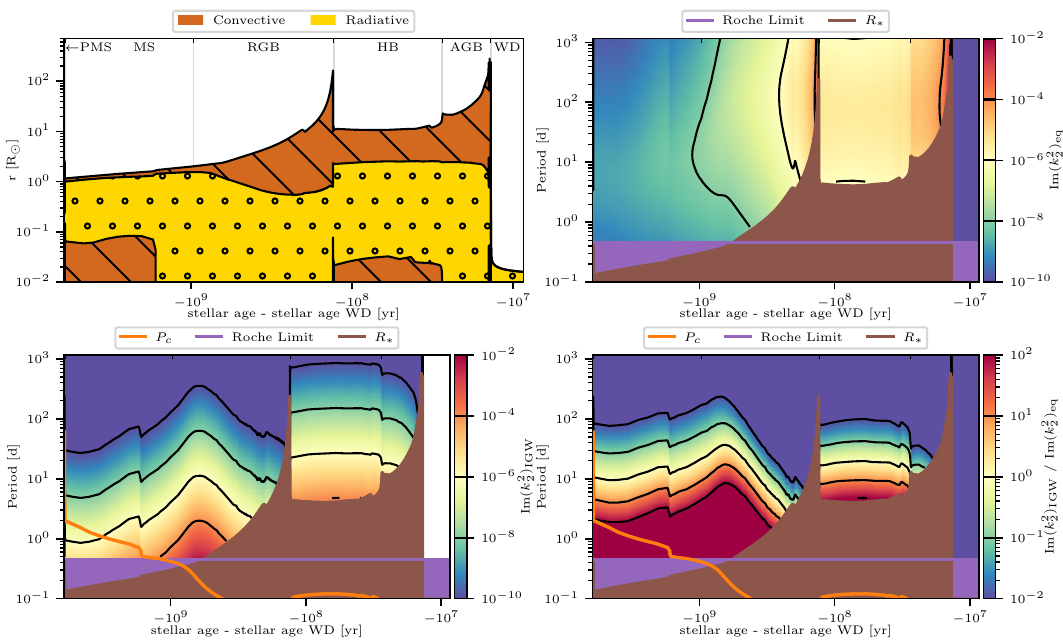}
        \caption{Same as Fig.~\ref{fig:FullM20}, but for a $M_\text{ZAMS}=1.2$ M$_\odot$ star.}\label{fig:FullM12}
    \end{figure*}
    \begin{figure*}
        \centering
        \includegraphics[width=\linewidth]{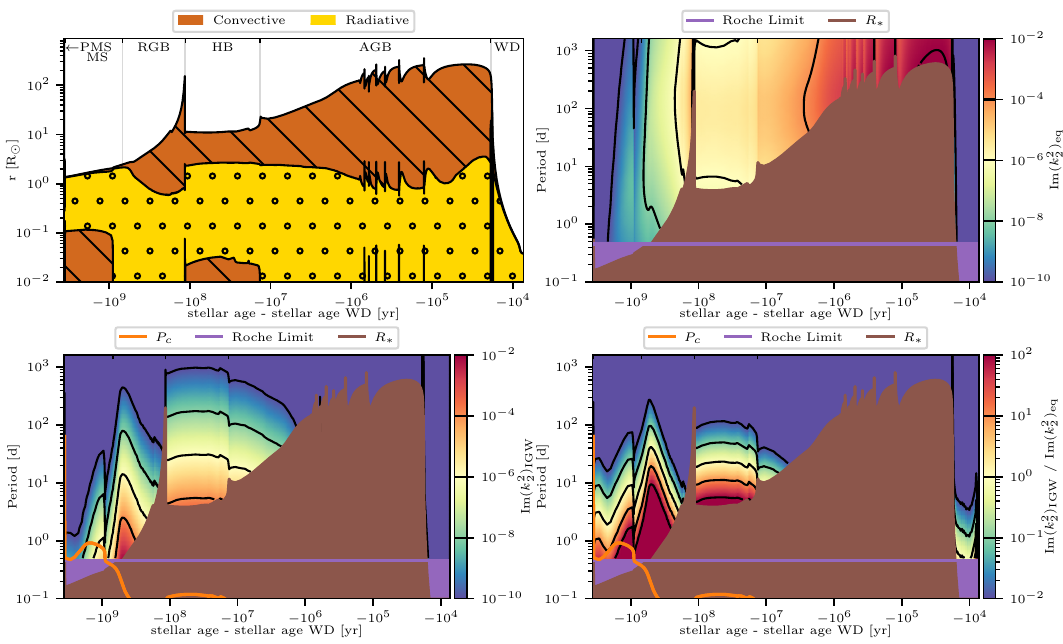}
        \caption{Same as Fig.~\ref{fig:FullM20}, but for a $M_\text{ZAMS}=1.4$ M$_\odot$ star.}\label{fig:FullM14}
    \end{figure*}
    \begin{figure*}
        \centering
        \includegraphics[width=\linewidth]{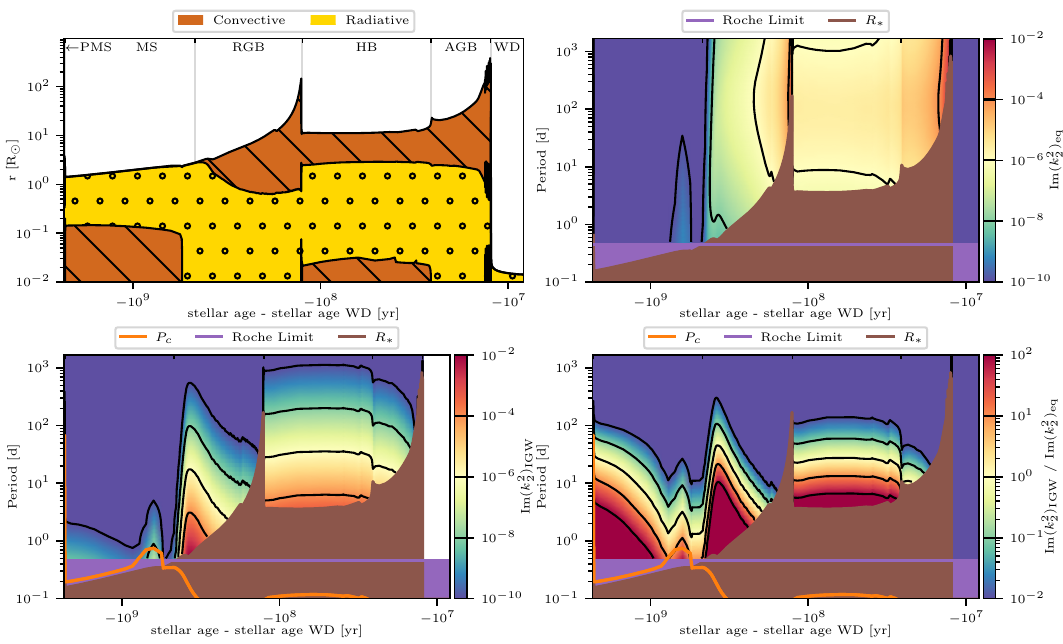}
        \caption{Same as Fig.~\ref{fig:FullM20}, but for a $M_\text{ZAMS}=1.6$ M$_\odot$ star.}\label{fig:FullM16}
    \end{figure*}
    \begin{figure*}
        \centering
        \includegraphics[width=\linewidth]{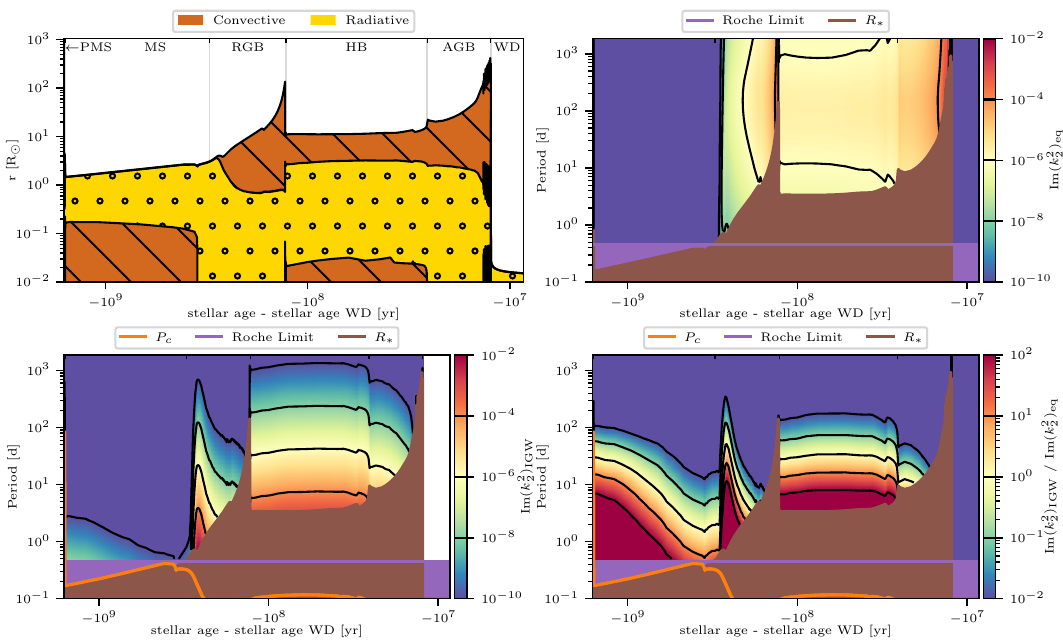}
        \caption{Same as Fig.~\ref{fig:FullM20}, but for a $M_\text{ZAMS}=1.8$ M$_\odot$ star.}\label{fig:FullM18}
    \end{figure*}
    \begin{figure*}
        \centering
        \includegraphics[width=\linewidth]{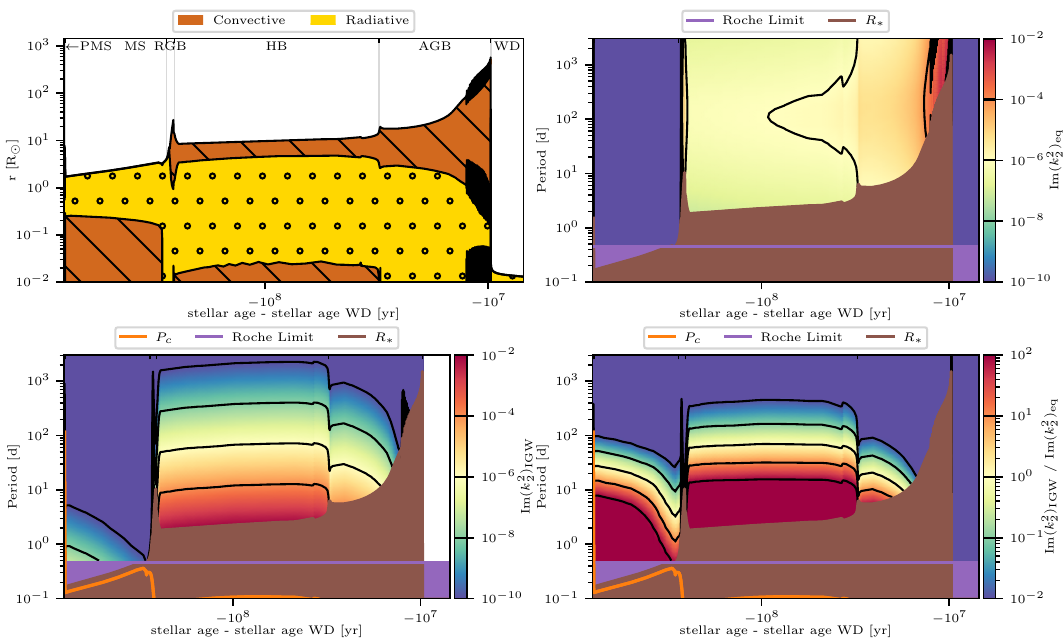}
        \caption{Same as Fig.~\ref{fig:FullM20}, but for a $M_\text{ZAMS}=2.5$ M$_\odot$ star.}\label{fig:FullM25}
    \end{figure*}
    \begin{figure*}
        \centering
        \includegraphics[width=\linewidth]{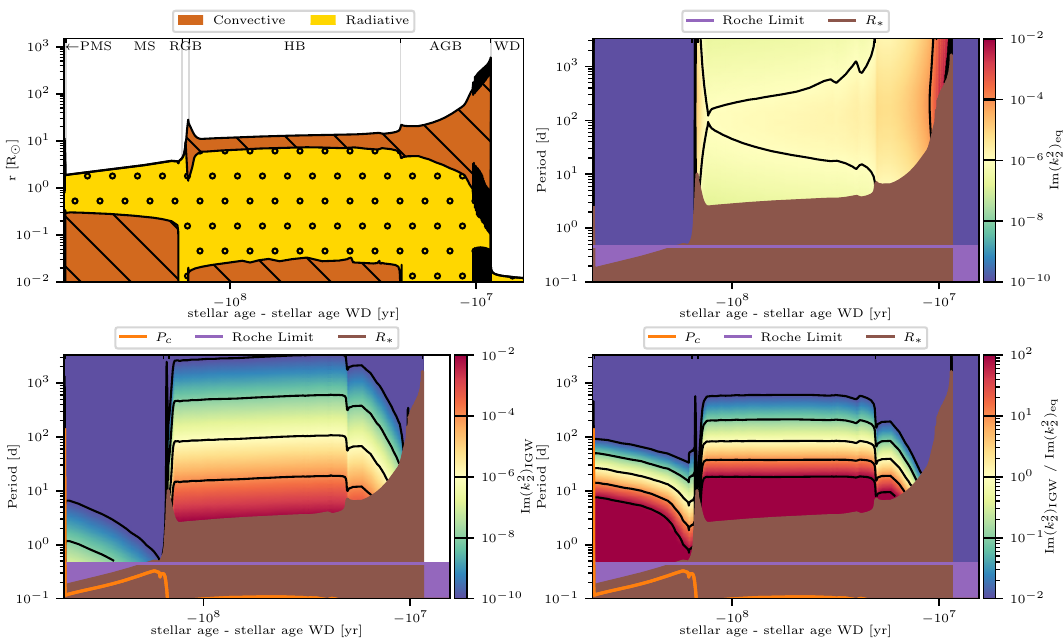}
        \caption{Same as Fig.~\ref{fig:FullM20}, but for a $M_\text{ZAMS}=3$ M$_\odot$ star.}\label{fig:FullM30}
    \end{figure*}

\clearpage
\section{MESA inlinst}\label{sec:Inlist}
    The MESA (release-r23.05.1) inlist we use in this study is reported hereafter. Here ``initial\_mass = mass'' refers to the initial mass of the star and ``Blocker\_scaling\_factor = Blocker'' to the Bl{\"o}cker scaling factor for AGB mass-loss; $\eta_\text{bl{\"o}cker}=0.05$ for masses below 2 $M_\odot$ and $\eta_\text{bl{\"o}cker}=0.1$ for masses above 2 $M_\odot$.

\begin{verbatim}
&star_job

    create_pre_main_sequence_model = .true.

    ! network
    auto_extend_net = .true.
    h_he_net = 'basic.net'
    co_net = 'co_burn.net'
    adv_net = 'approx21.net'

    ! opacities
    initial_zfracs = 6 ! for L03 solar scaling

/ !end of star_job namelist


&kap

    Zbase = 0.0134d0

    use_Type2_opacities = .true.
    kap_file_prefix = 'a09'
    kap_CO_prefix = 'a09_co'
    kap_lowT_prefix = 'AESOPUS'
    AESOPUS_filename = 'AESOPUS_AGSS09.h5'

/ ! end of kap namelist


&controls

    log_L_lower_limit = -1

    varcontrol_target = 1d-4
    min_timestep_limit = 1d-10

    initial_mass = Mass
    initial_z = 0.0134d0

    mixing_length_alpha = 1.931
    MLT_option = 'Henyey'

    atm_option = 'T_tau'
    atm_T_tau_relation = 'Eddington'
    atm_T_tau_opacity = 'fixed'

    cool_wind_RGB_scheme = 'Reimers'
    cool_wind_AGB_scheme = 'Blocker'
    RGB_to_AGB_wind_switch = 1d-4
    Reimers_scaling_factor = 0.477d0
    Blocker_scaling_factor = Blocker

    set_min_D_mix = .true.
    min_D_mix = 1d1

    num_cells_for_smooth_brunt_B = 10
    threshold_for_smooth_brunt_B = 0.1
    num_cells_for_smooth_gradL
            _composition_term = 10
    threshold_for_smooth_gradL
            _composition_term = 0.02

    report_solver_progress = .true.
    use_gold_tolerances = .false.

    solver_iters_timestep_limit = 20
    solver_max_tries_before_reject = 30

    energy_eqn_option = 'dedt'
    convergence_ignore_equL_residuals = .true.
    corr_coeff_limit = 1d-1
    max_abs_rel_run_E_err = 1d-1
    warn_when_large_rel_run_E_err = 1d-3
    ignore_too_large_correction = .true.
    scale_max_correction = 1d-2
    ignore_min_corr_coeff_for
        _scale_max_correction = .true.
    ignore_species_in_max_correction = .true.
    use_superad_reduction = .true.

    log_directory = 'LOGS'
    photo_interval = 10
    max_num_profile_models = 100000
    profile_interval = 10
    history_interval = 1
    terminal_interval = 1
    write_header_frequency = 1

/ ! end of controls namelist
\end{verbatim}

\end{appendix}

\end{document}